\documentclass[english]{iopart}

\usepackage{graphicx}
\usepackage{amssymb}

\usepackage{url}

\newcommand{\ann}{\mathrm{ann}}

\newcounter{fig}
\begin{document}

\title[Lattice Green Functions]{\Large Lattice Green Functions: the 
        $d$-dimensional face-centred cubic lattice,
$d=\,8,\, 9,\, 10,\, 11,\, 12$}

\author{S. Hassani$^\S$, Ch. Koutschan$^\dagger$, J-M. Maillard$^\pounds$, N. Zenine$^\S$  
}
\address{\S  Centre de Recherche Nucl\'eaire d'Alger, 
2 Bd. Frantz Fanon, BP 399, 16000 Alger, Algeria}
\address{$^\dagger$ Johann Radon Institute for Computational and Applied
Mathematics, Austrian Academy of Sciences, Altenberger Stra\ss e 69, 
A-4040 Linz, Austria} 
\address{$^\pounds$ LPTMC, UMR 7600 CNRS, 
Universit\'e de Paris, Tour 23,  5\`eme \'etage, case 121, 
 4 Place Jussieu, 75252 Paris Cedex 05, France}

\vskip .2cm 

E-mail:  christoph.koutschan@ricam.oeaw.ac.at, maillard@lptmc.jussieu.fr,
\vskip .01cm 
 njzenine@yahoo.com

\vskip .3cm 

\vskip .2cm 

{\em Dedicated to A. J. Guttmann, for his 70th birthday.}

\begin{abstract}
We previously reported on a recursive method to generate the expansion 
of the lattice Green function of the $d$-dimensional face-centred cubic
lattice (fcc). The method was used to generate many coefficients for $\,d=\,7$
and the corresponding linear differential equation has been obtained.
In this paper, we show the strength and the limit of the method by 
producing the series and the corresponding linear differential equations 
for $\,d=\, 8, \,9,\, 10,\, 11,\, 12$.
The differential Galois groups of these linear differential equations 
are shown to be symplectic for $\,d=\,8, \,10, \,12$
 and orthogonal for $\,d=\,9, \,11$. 
The recursion relation naturally provides a $2$-dimensional array 
$\, t_d(n,j)$ where only the coefficients $t_d(n,0)$ correspond to the 
coefficients of the lattice Green function of the $\, d$-dimensional fcc. 
The coefficients $\, t_d(n,j)$ are associated to $\, D$-finite bivariate series
annihilated by linear partial differential equations that we analyze.

\vskip .3cm

\noindent {\bf PACS}:
05.50.+q, 05.10.-a, 02.10.De, 02.10.Ox


\vskip .3cm

\noindent {\bf Key-words}:
Lattice Green function, face-centred cubic lattice, long series expansions,
partial differential equations, Fuchsian linear differential equations, 
differential Galois groups, D-finite systems, 
indicial exponents, apparent sing\-ularities, Landau conditions.

\end{abstract} 

\vskip .3cm

\section{Introduction}
\label{Intro}

The \emph{lattice Green function} (LGF) of the $d$-dimensional 
face-centred cubic (fcc) lattice is given by a $d$-fold integral
whose expansion around the origin is hard to obtain as the dimension goes 
higher~\cite{guttmann-2010,broadhurst-2009,guttmann-1993,koutschan-2013}.
Only for $d=3$ a closed form is known~\cite{Joyce}. But
since the integrand is of a very simple form---a rational function, after an
appropriate variable transform---it follows from the theory of holonomic
functions~\cite{Lipshitz, zeilberger-1990} that those integrals are \emph{D-finite}
or \emph{holonomic}, i.e. each of them satisfies a linear ordinary 
differential equation (ODE) with polynomial coefficients.
For $d=4$ the corresponding linear ODE was obtained in~\cite{guttmann-2009},
for $d=5$ in~\cite{broadhurst-2009}, and for $d=6$ in~\cite{koutschan-2013},
by different methods.
In a previous paper~\cite{2015-LGF-fcc7} we forwarded a {\em recursive method}
that was efficient enough to allow us generate many series coefficients 
for $d=\, 7$ necessary to obtain the linear ODE.
Since the recursion parameter in the method is the dimension $\, d$, we have
obtained many short series for $d$ as high as~$45$.
From these data and the {\em Landau equations method}~\cite{2007-PhiH-integrals}
 on the integrals, we inferred 
many properties that we conjecture to be common to all the linear 
ODEs of $d$-dimensional fcc lattices.
The order-eleven linear differential operator, corresponding to the linear 
differential equation, we have obtained~\cite{2015-LGF-fcc7} for the 
lattice Green function of the $7$-dimensional face-centred cubic (fcc) 
has been found to verify a property recently forwarded~\cite{2015-Canonical-Special}.
This property is a canonical decomposition of irreducible linear differential
operators with symplectic or orthogonal differential Galois groups and corresponds
to the occurrence of a homomorphism of the operator and its adjoint.
This property has been seen to 
occur~\cite{2014-DiffAlg-LGFCY,2015-Scaling-Ising,2014-SpecGeom-Ising}
for many linear differential operators that emerge in lattice statistical physics 
and enumerative combinatorics. 

\vskip 0.1cm

In this paper, we show the strength and the limit of the method.
With some technical improvements in the computations, we show how much high 
in dimension~$d$ we can go in generating
sufficiently many terms of the LGF series in order to obtain the
corresponding linear ODE. We find that the conjectures (especially 
on the singularities) given in~\cite{2015-LGF-fcc7} are all verified.
We also find that the canonical decompositions of the operators follow 
the scheme given in~\cite{2015-Canonical-Special}.

Furthermore, the recursion relation gives a $2$-dimensional array 
$\, t_d(n,j)$ where only the coefficients $\, t_d(n,0)$ correspond 
to the coefficients of the lattice Green function of the 
$\, d$-dimensional fcc. We give the integrals whose expansion
gives {\em bivariate series} with coefficients $\, t_d(n,j)$,
 and address the \emph{D-finite systems} 
that annihilate these bivariate series.

\vskip 0.1cm

The paper is organized as follows. Recalls are given in Section~\ref{sec:gen} 
where improvements of the method and some computational details are 
also given. Section~\ref{sec:odes} deals with our results on the 
differential equations annihilating the LGF of the
 $ \, d$-dimensional fcc lattice for $\, d= \, 8,\,  9, \, 10,\,  11,\,  12$.
The orders and singularities of all these linear ODEs
are seen in agreement with our conjectures 
and computed {\em Landau singularities} in~\cite{2015-LGF-fcc7}.
In Section~\ref{sec:galois}, we show that the 
{\em differential Galois groups} of the operators are 
symplectic for $\, d=\, 8, \,d=\, 10, \,d=\, 12$ and orthogonal 
for $d=\, 9, \,d=\, 11$. We give for the
operators corresponding to $d=\, 8$ and $d=\, 9$ the canonical 
decomposition that should~\cite{2015-Canonical-Special} occur 
for the operators with symplectic or orthogonal differential 
Galois groups. 
In Section~\ref{sec:coupling} we give the $d-$dimensional integrals
depending on two variables $(z, y)$ whose expansion around $(0, 0)$ (and
integration) writes in terms of the coefficients $\, t_d(n,j)$.
These coefficients generate a $ \, D$-finite bivariate series $\,T_d(z, y)$ 
annihilated by a system of partial differential equations (PDE)
that we give in Section~\ref{partialdiff} for the case $d=2$.
Switching to a system of {\em decoupled} PDE, we give in Section~\ref{ODET2}
the solutions that combine to match  $\,T_2(z, y)$.
Section~\ref{T3} deals with $d=3$, where we focus on one solution which,
remarkably, is a modular form.
We present some remarks in Section~\ref{remcom} and conclude in
Section~\ref{Conclusion}.

\section{LGF generation of series}
\label{sec:gen}

\subsection{Recalls on the recursive method}
The lattice Green function (LGF) of the $d$-dimensional face-centred 
cubic (fcc) lattice reads
\begin{eqnarray}
\label{LGFd}
\hspace{-0.95in}&& \quad \qquad   \qquad     
LGF_d(x) \,\, =\,\,\, \, 
{\frac{1}{\pi^d}}\, \int_0^\pi \cdots \,
 \int_0^\pi \, {\frac{\mathrm{d}k_1\, 
 \cdots \,\,  \mathrm{d}k_d}{1\, -x \cdot \, \lambda_d}}, 
\end{eqnarray}
where $\lambda_d$, called the structure function of the lattice, is given by:
\begin{eqnarray}
\hspace{-0.95in}&& \quad \qquad   \qquad     
\lambda_d \,\, =\,\,\, { d \choose 2}^{-1} \cdot \, \sum_{i=1}^d \, \sum_{j=i+1}^d 
\cos(k_i) \cdot \, \cos(k_j).
\end{eqnarray}
The expansion around the origin of $\, LGF_d(x)$ is given by
\begin{eqnarray}
\hspace{-0.95in}&& \qquad \quad  \qquad  
LGF_d(z) \,\,\,   =\, \, \, \, 
  \sum_{n=0} \, z^n \cdot \, t_d(n,0), \qquad  
z \,\,  = \, \, \, {x \over 4}  \cdot \, { d \choose 2}^{-1}, 
\end{eqnarray}
where the array\footnote{
See Section~\ref{sec:coupling} for the integral representation 
corresponding to $\,t_d(n,j)$.} $\,t_d(n,j)$ is obtained with 
the recursive relation (see Section 2 in~\cite{2015-LGF-fcc7})
\begin{eqnarray}
\label{Tdnj}
\hspace{-0.95in} \,\,   t_d (n, j) \, \,=\,\, \, \, 
 \sum_{p=0}^n \, \sum_{q=q_1}^{q_2} \,
{ n \choose p} \, { 2 j \choose 2 q+p-n } \,
 { 2 n\, +2 j\, -2 p\, -2 q \choose n+j-p-q} \cdot \, t_{d-1} (p, q), 
\\
\fl \qquad \quad \quad q_1\,  =\,\,   [(n-p+1)/2], 
\quad \quad \quad \quad  q_2 \, = \,\,  [(n-p+2 j)/2], 
\end{eqnarray}
where $[x]$ is the integer part of $x$.
To start the recursion, one needs: 
\begin{eqnarray}
\label{T2nj}
\hspace{-0.95in} && \qquad\qquad \, \, 
 t_2 (n, j)\,  \,=\,\, \, \, \sum_{p=p_1}^{p_2} \,
{ 2 p  \choose p} \, { 2 j \choose 2 p-n } \, { 2 n+2 j-2 p \choose n+j-p},
 \\
\label{T2nj2}
\hspace{-0.95in} && \qquad  \qquad \qquad \quad \quad 
p_1 \, = \,\,  [(n+1)/2],
 \quad \quad \quad  \quad  \quad   p_2 \, =\, \,  [(n+2 j)/2].
\end{eqnarray}

\subsection{Computational details}
\label{computdetails}

The shape of the recurrence~(\ref{Tdnj}) suggests to start with the
two-dimensional array $\, t_2(n,j)$, then compute $\, t_3(n,j)$, and
 so on. Once $\, t_3(n,j)$ is completed, the data of $\, t_2(n,j)$ is not 
any more needed since the recurrence~(\ref{Tdnj}) is of 
order~$1$ with respect to~$d$. Also from the
recurrence it is easy to see that for 
computing $\, t_d(n,0)$, $\, 0\, \leq n \, \leq\,  N$,
the desired coefficients of the Taylor series, one needs the values of
$\, t_{d-1}(n,j)$ for 
$ \,0 \,\leq  \,n \,\leq \, N$ and $ \,0 \,\leq \, j \,\leq[(N-n)/2]$ 
(the same range applies to the arrays $t_{d-2},\dots, \, t_2$). The 
advantage of such an implementation is that 
{\em it stores only $ \,O(N^2)$ elements, which are
integers}. The disadvantage is that one has to fix $\, N$ at the 
very beginning, but the number of terms needed for constructing the 
linear differential operator is {\em not known in advance}.

If one looks at the recurrence~(\ref{Tdnj}) more closely, one discovers the
remarkable fact that neither its coefficients, i.e. the product of the three
binomials, nor its support, i.e. the summation bounds, depend on the
parameter~$d$.  Hence, in the previous approach, the coefficients are the same
in each step, but they are recomputed in each iteration $\,d= \,3,\,4,\,\dots$, 
which is clearly a waste of computational resources. In principle, 
we could collect all the coefficients in a big matrix~$A$ 
that maps the array $\, t_{d-1}$ to~$t_d$, so that 
$ \,t_d= \,A^{d-2} \cdot  \,t_2$. For this purpose 
the two-dimensional arrays $\, t_d(n,j)$, 
$ \,0 \,\leq \, n \,\leq \, N$, $ \,0 \,\leq \, j \, \leq[(N-n)/2]$, 
have to be
represented as vectors of dimension $\,[(N/2+1)^2]$. So the whole computation
then boils down to compute the power of some matrix, and then multiply it to
the vector that corresponds to~$t_2$.  The problem is that the matrix~$A$ has
dimension $[(N/2+1)^2]\times[(N/2+1)^2]$, which already for the
$8$-dimensional fcc lattice (where we need at least $N=\,704$ Taylor
coefficients, see Table~\ref{Ta:1}) means a $\,124609 \times 124609$ square
matrix. Of course, $A$ is not dense. A simple calculation reveals that it has
$(N+2)(N+4)(N^2+4N+12)/96$ nonzero entries if $\,N$ is even, and
$(N+1)(N+3)(N^2+6N+17)/96$ nonzero entries when $\,N$ is odd. It follows that
$A$ has sparsity $\,1/6$. Nevertheless it would require a considerable
and impractical amount of memory to store the full matrix: for $d=8$ it has
about $2.6$ billion nonzero entries (which themselves are big integers), and
for $d=\,11$, where we need $N=\,2464$ terms, it has about $386$ billion 
nonzero entries.

From the above discussion we are led to the following considerations: on the
one hand, we would like to avoid recomputation of the coefficients, and on the
other hand, we do not want to compute them all at once. Moreover, it is
desirable to have a program that computes the Taylor coefficients one after
the other, so that one does not have to fix $\,N$ at the very beginning. The
following algorithm satisfies all three requirements. The main loop is
$\,n=\,0,\,1,\,2,\,\dots$ and in each iteration the values
$\,t_d(0,n/2),\, t_d(2,n/2-1), \, t_d(4,n/2-2),\,\dots,\, t_d(n,0)$ if $\, n$ 
is even (resp. $\,t_d(1,(n-1)/2),\, t_d(3,(n-3)/2),\, \dots, \, t_d(n,0)$ 
for odd~$n$) are computed in the given order, for all $ \, d$ 
between $ \, 2$ and the dimension of the lattice. For sake of brevity, 
and without loss of generality, we will focus on
the case of even~$n$ in the following.
 Note that in this way all the data
that is required for $\, t_d(2k,n/2-k)$ is already available. Similarly 
as before, we can obtain $\, t_d(2k,n/2-k)$ as the 
scalar product $\,a \cdot \, t_{d-1}$, where
$\,a$ is a row vector and the two-dimensional array $\, t_{d-1}$, again, 
has to be interpreted as a single column vector. The vector~$a$ corresponds 
to a single row of the above-mentioned matrix~$A$.  Then
$\,t_3(2k,n/2-k),\, t_4(2k,n/2-k),\,\dots$ are computed by using always 
the same vector~$a$, so that any recomputation of coefficients is 
avoided. The only drawback of this approach is that one has to keep 
the whole three-dimensional array $\, t_d(n,j)$ in memory, and therefore 
this method is {\em more memory-intensive} than the naive approach 
(by a factor of approximately $\,d/2$).

The described computational scheme allows for lots of further (technical) 
improvements, some of which we want to mention briefly here. For example, 
{\em one does not need to compute} the vector~$a$ from scratch for 
each~$k$, but reuse the previous one by adding and deleting a few 
entries, and apply the simple recurrences
for binomial coefficients:
\begin{eqnarray}
\label{bin}
\hspace{-0.95in} && \quad  \, \, 
  {n+1\choose k}\,  = \,\, \, \frac{n+1}{n-k+1} \cdot  \, {n\choose k}
  \quad \, \, \,  \,\, \,  \, \mathrm{and} \quad \quad \,  \, 
  {n\choose k+1} \,  =  \,\, \,  \frac{n-k}{k+1} \cdot  \, {n\choose k}.
\end{eqnarray}
Multiplying by a simple rational number is much cheaper than calculating a
binomial coefficient.

With a little effort there is also the possibility to parallelize the
computation. This can be done by splitting the sequence
$ \, t_d(0,n/2), \, t_d(2,n/2-1), \, \dots, \, t_d(n,0)$ 
into parts, each of which is done by a
single processor. The only caveat is the contribution of
$\,t_d(0,n/2),\,\dots, \, t_d(2k-2,n/2-k+1)$ to the 
computation of $ \, t_d(2k,n/2-k)$,
which has to be postponed until all processors have finished their task. This
causes some synchronization overhead at each iteration of~$n$, which prevents
us from using an excessive amount of processors. For example, the computation
time for the required $N= \, 999$ terms for the $9$-dimensional fcc lattice
dropped from $60$ hours to $7.5$ hours by using $10$ parallel processors.

For the interested reader we also mention the timings for the other dimensions
of the lattice considered here: in the $10$-dimensional case we obtained the
necessary $N=\, 1739$ terms in $3$ days using 20 parallel 
processes, for $ \, d= \, 11$ the same number of processes was running for $18$ 
days to compute the $2464$ terms mentioned in Table~\ref{Ta:1}. 
For the $12$-dimensional fcc lattice we only computed modulo $p= \, 2^{31}-1$, 
and found that the minimal number of terms necessary for constructing 
the linear differential operator is $3618$: these were obtained in $10$ days 
using $25$ parallel processes.

\begin{table}[htdp]
\caption{
 The number of terms $\, N_m$ (and $\, N_0$) needed 
to obtain the minimal-order linear ODE of order $ \, Q_{min}$ 
(and the optimal-order linear ODE of order $\, Q_{opt}$)
annihilating $\, LGF_d(x)$.
 }

\label{Ta:1}
\begin{center}
\begin{tabular}{|c|c|c|c|c|}\hline
    $d$ &  $N_m$ &  $N_0$ & $N_m-N_0$ & $Q_{opt}-Q_{min}$       \\ \hline 
\hline
    4 & 40  & 40   & 0 & 4-4\,=\,\,0     \\
    5 & 98  & 88   & 10 & 7-6\,=\,\,1     \\
    6 & 342  &  228  & 114 & 11-8\,= \,\, 3    \\
    7 & 732  &  391  & 341 & 16-11\,=\,\,5     \\
    8 & 1740  & 704   & 1036 & 21-14\,=\,\,7     \\
    9 & 2964  & 999   & 1965 & 26-18\,=\,\,8     \\
   10 & 6509  & 1739   & 4770 & 36-22\,=\,\,14     \\
   11 & 10864 & 2464   & 8400 & 43-27\,=\,\,16    \\
   12 & 19503 & 3618   & 15885 & 53-32\,=\,\, 21    \\
\hline
 \end{tabular}
\end{center}
\end{table}

\section{The differential equations of the LGF 
 of the $\, d$-dimensional fcc lattice, $\, d=\,8, \cdots, 12$}
\label{sec:odes}

We obtain the corresponding linear ODE using an ansatz
$\sum_{i,j}c_{i,j}x^j \, \left(x \, {d \over dx} \right)^i$
with undetermined coefficients
$c_{i,j}\in\mathbb{Q}$. Substituting the generated series into this ansatz and
equating the coefficients with respect to~$x$ to zero yields a linear system for
the~$c_{i,j}$. The result is very trustworthy once we use a sufficient amount
of series data, i.e. such that the resulting linear system has more equations
than unknowns. In the computer algebra literature this methodology is referred
to as \emph{guessing}, which somehow hides the fact that it is a completely
algorithmic and constructive method.  The linear ODE for
$\,d=\,8,\, 9, \,10,\, 11$ are obtained in exact arithmetic and the linear ODE
for $\,d=\,12$ is obtained modulo one prime.  These linear ODEs are given in
electronic form in~\cite{koutschan-webpage}.

\vskip 0.1cm

The linear differential operators 
corresponding to $\,d=\,8,\, 9, \,10,\, 11,\, 12$ are called, respectively, 
$\,G_{14}^{8Dfcc}$, $\,G_{18}^{9Dfcc}$, $\,G_{22}^{10Dfcc}$,  $\,G_{27}^{11Dfcc}$,
and $\,G_{32}^{12Dfcc}$,
 where the subscript refers to the order of the linear ODE.
These orders are in agreement with the conjecture given 
in~\cite{2015-LGF-fcc7}:
\begin{eqnarray}
\label{orderq}
\hspace{-0.75in} && \quad \quad \quad \quad \qquad \,\, 
q \, \,=\,\,\, \, \,  {\frac{d^2}{4}} \,\,\, 
   - {\frac{d}{2}}\, \,\,  \, + {\frac{17}{8}}
 \,\,\,  - {\frac{(-)^d}{8}}. 
\end{eqnarray}

\vskip 0.1cm

We have also agreement with the regular singularities 
$\, x_s$ obtained by the {\em Landau equations method}, 
which read~\cite{2015-LGF-fcc7}
\begin{eqnarray}
\label{thexs}
 \hspace{-0.75in} && \quad \quad \quad \qquad \qquad
  x_s \,\,  = \,\, \, \, 
 { d \choose 2} \cdot \, {\frac{1}{ \xi(d, k, j)}}, 
\end{eqnarray}
where
\begin{eqnarray}
\label{singfirstset}
\hspace{-0.95in} && \quad \quad  \quad
\xi(d, k, j)\, \,  \,=\, \, \, \, \,
{\frac{{d}^{2}\, \, 
- \, (k \, +4\,j\, +1)  \cdot\,  d\,\,\,
  +4\,{j}^{2}\, +k\,+4\,j\,k }{2 \cdot \, \,(1-k) }},
 \\
\hspace{-0.95in} &&  \qquad 
 \quad \quad \, \, \,    {\rm with} \quad  \quad \quad  \quad 
k=\,\, 0,\,  2, \, 3,\,  \cdots, \,  d-1, 
\quad  \quad  \quad  j= \, 0, \, \cdots,\,  [(d-k)/2],  
\nonumber
\end{eqnarray}
and where $\, [x]$ is the integer part of $\, x$. 

The singularities as they occur in front of the head derivative of respectively 
$\,G_{14}^{8Dfcc}$, $\,G_{18}^{9Dfcc}$, $\,G_{22}^{10Dfcc}$, $\,G_{27}^{11Dfcc}$, 
$\,G_{32}^{12Dfcc}$ 
are given in \ref{sing891011}.

\vskip 0.1cm

As far as the local exponents at the singularities are concerned, one remarks
that the regular pattern seen~\cite{2015-LGF-fcc7}  at $\, x=\,0$ continues.
Note that this is the pattern from which we inferred the order of the linear ODE.
The local exponents, at each singularity, are given in Table~\ref{Ta:2}.
The singularities $\,x_d$, which read 
$\,x_3 = -3$, $\,x_5 = -5$, $\,x_6 = -15$, $\,x_7 = -7$, $\,x_8 = -14$, $\,x_9 = -9$,
$\,x_{10}=-15$, $\,x_{11}=-11$, $\,x_{12}=-33/2$, seem to be given by:
\begin{eqnarray}
\label{seem}
\hspace{-0.95in} &&  \qquad \quad \qquad 
\quad   x_d \, \,=\, \,\,
-{\frac {2  \, d \cdot \, (d-1) }{2\,d \,  -5 \, \, -3 \cdot \, (-1)^{d}}}.
\end{eqnarray}

For $d= \,   11$, there is also the singularity $ \, x= \, -55$ not included in 
``others'' (not shown in Table \ref{Ta:2}) with local exponents $\, 9/2,\,  11/2$.

\begin{table}[htdp]
\caption{
The local exponents at the regular singularities. Only the exponents
giving a singular behavior are shown.
}
\label{Ta:2}
\begin{center}
\begin{tabular}{|c|c|c|c|c|c|}\hline
    $d$ &  $x=0$ &  $x=\infty$ & $x=1$       & $x_d$ &  others         \\ \hline 
\hline
    3 & $0^3$          &  $3/2$    &  $1/2$  &  $0^3$          &  \\
    4 & $0^4$          &  $2^2$    &  $1^2$  &                 &  $1^2$  \\
    5 & $0^5, 1$       &  $5/2$    &  $3/2$  & $3/4, 5/4$      &  $3/2$  \\
    6 & $0^6, 1^2$     &  $3^2$    &  $2^2$  & $2^2, 3^2$      &  $2^2$  \\
    7 & $0^7, 1^3, 2$  &  $7/2$    &  $5/2$  & $3/2, 5/2, 2^3$ &  $5/2$ \\
    8 & $0^8, 1^4, 2^2$  &  $4^2$    &  $3^2$  & $3^2, 4^2$ & $3^2$  \\
    9 & $0^9, 1^5, 2^3, 3$  & $9/2, 11/2$     & $7/2$   & $9/4, 11/4, 13/4, 15/4$ & $7/2$  \\
   10 & $0^{10}, 1^6, 2^4, 3^2$  & $5^2$           & $4^2$ &  $4^2, 5^2$ &  $4^2$  \\
   11 & $0^{11}, 1^7, 2^5, 3^3, 4$ & $11/2$   & $9/2$   &  $7/2, 9/2, 3^2, 4^3, 5^2$ &  $9/2$  \\
   12 & $0^{12}, 1^8, 2^6, 3^4, 4^2$  &   $6^2$      &   $5^2$      & $5^2, 6^2$   & $5^2$ \\
\hline
 \end{tabular}
\end{center}
\end{table}

\section{The differential Galois groups of $\, G_{q}^{dDfcc}$, $\,d=\,8,\,  9, \,10,\, 11,\, 12$}
\label{sec:galois}

The equivalence of two properties, namely 
the {\em homomorphism of the operator with its adjoint}, and either the occurrence 
of a {\em rational solution} for the {\em symmetric (or exterior)} square 
of that operator, or the  drop of order of these squares,
have been seen for many linear differential operators~\cite{2014-DiffAlg-LGFCY}.
The operators with these properties are such that their {\em differential Galois 
groups} are included in the {\em symplectic or orthogonal differential groups}. 
We have also shown that such operators have a 
``canonical decomposition''~\cite{2015-Canonical-Special},
which means that they can be written in terms 
of ``tower of intertwiners''.
These properties hold also for the (non-Fuchsian) operators emerging in the 
square Ising model at the scaling limit~\cite{2015-Scaling-Ising}.

\vskip 0.1cm

For the linear differential operators annihilating
 $\, LGF_d(x)$, ($d=\,5, \,6, \,7$), 
of the fcc lattice, these properties hold~\cite{2015-LGF-fcc7, 2014-DiffAlg-LGFCY}.
For instance, the order-eleven operator  $\,G_{11}^{7Dfcc}$ (corresponding 
to $\,LFG_7(x)$) has the following canonical decomposition~\cite{2015-LGF-fcc7}
\begin{eqnarray}
\label{canon-decompL11}
\hspace{-0.95in}&& \,   \quad    \quad   
G_{11}^{7Dfcc} \,   = \, \, \,\,
  (A_{1} \cdot \,B_{1} \cdot \, C_{1} \cdot \, D_{1} \cdot \,  E_{7}
\,\,\, + A_{1} \cdot \, B_{1} \cdot \, E_{7} \,  \,
+ \, A_{1} \cdot \, D_{1} \cdot \,  E_{7} \, 
\nonumber \\
\label{last3}
\hspace{-0.95in}&&  \quad \quad   \quad  \,  \quad \quad    \quad 
 + \, A_{1} \cdot \, B_{1} \cdot \,  C_{1}
+ \, C_{1} \cdot \, D_{1} \cdot \, E_{7} 
\,\,\,\,  + E_{7} \, \, + \, C_{1}\,+ \, A_{1}) \cdot r(x),  
\end{eqnarray}
where $\, r(x)$ is a rational function, and the 
factors (the indices correspond to their orders) 
are all {\em self-adjoint} linear differential operators.

The decomposition (\ref{canon-decompL11}) occurs because $\, G_{11}^{7Dfcc}$ 
is {\em non-trivially homomorphic to its adjoint}, and 
the decomposition is obtained through a sequence of Euclidean 
right divisions (see Section~5 in~\cite{2015-LGF-fcc7}).

From this decomposition one understands easily why the {\em symmetric square} 
of $\,G_{11}^{7Dfcc}$ is of order $\, 65$, instead of the generically expected 
order $\, 66$.  The symmetric square of the self-adjoint order-seven 
linear differential operator $\, E_{7}$ is of order $27$,
instead of the generically expected order $28$
(see~\cite{2015-LGF-fcc7, 2015-Canonical-Special} for details).
From the decomposition (\ref{canon-decompL11}) one immediately deduces
the decomposition of the adjoint of $\,G_{11}^{7Dfcc}$ (because the factors 
are self-adjoints).
The symmetric square of the adjoint $\,G_{11}^{7Dfcc}$ will annihilate a
rational solution which is the square of the solution 
of the order-one operator $A_1$.
The differential Galois group of $\,G_{11}^{7Dfcc}$ is included in 
$\, SO(11, \, \mathbb{C})$.

\vskip 0.1cm

In the sequel, we show that the operators $ \, G_{14}^{8Dfcc}$, 
$ \, G_{18}^{9Dfcc}$, $ \, G_{22}^{10Dfcc}$, $\, G_{27}^{11Dfcc}$ 
and $\,G_{32}^{12Dfcc}$
verify the same properties as the operators $ \, G_{11}^{7Dfcc}$ 
(and $\, G_{6}^{5Dfcc}$ for $\, d= \, 5$, see~\cite{broadhurst-2009},
 $\, G_{8}^{6Dfcc}$ for $\, d=\, 6$, see~\cite{koutschan-2013}).
 For $\, d$ odd (resp. $d$ even), one must consider the 
symmetric square (resp. exterior square) of the operator.

\vskip 0.1cm

Note that to compute the homomorphism, for our purpose, between an operator
and its adjoint, the operator should be irreducible (see Section 2.1 in
\cite{2014-DiffAlg-LGFCY}). We have shown in \cite{2015-LGF-fcc7} that 
$\, G_{6}^{5Dfcc}$, $\, G_{8}^{6Dfcc}$, $ \, G_{11}^{7Dfcc}$ are irreducible.
In the next section, we will assume that $ \, G_{14}^{8Dfcc}$, 
$ \, G_{18}^{9Dfcc}$, $ \, G_{22}^{10Dfcc}$, $\, G_{27}^{11Dfcc}$ 
and $\,G_{32}^{12Dfcc}$ are irreducible\footnote{The results given in 
(\ref{homoG14d8}) and (\ref{adjG18}) show that $ \, G_{14}^{8Dfcc}$
and $ \, G_{18}^{9Dfcc}$ are, indeed, irreducible.}.

\subsection{The differential Galois group of $\, G_{14}^{8Dfcc}$}
\label{8Dfcc}

Producing the 14 formal solutions of the linear differential operator 
$\, G_{14}^{8Dfcc}$, it is easy to show that its {\em exterior square} 
is of order $\,  90$, instead of the generically expected order $\, 91$. 
The differential Galois group of the operator $\, G_{14}^{8Dfcc}$ is 
included in $\, Sp(14, \,\mathbb{C})$.

The exterior square of the adjoint of $\, G_{14}^{8Dfcc}$ either has the
order $\,  90$, or annihilates a rational solution.
We find that the exterior square of the adjoint of $ \, G_{14}^{8Dfcc}$ 
annihilates the following rational function
\begin{eqnarray}
\label{RatSold8}
\hspace{-0.75in} && \quad \quad  \quad \, 
sol_R \, (ext^2 \, (adjoint (G_{14}^{8Dfcc})))
\,  \, =\, \, \, \,
{\frac{x^{21}  \cdot \, P_{84}(x)  \cdot \, S_8(x)^2}{ P_{14}(x)}}, 
\end{eqnarray}
where $ \, P_{14}(x)$ is the degree-$95$ {\em apparent polynomial} of 
$\, G_{14}^{8Dfcc}$, $ \, S_8(x)$ is a degree-8 polynomial corresponding to 
the finite singularities given in \ref{sing891011}, and 
$ \, P_{84}$ is a degree-84 polynomial.

From these results, we should expect, in the ``canonical decomposition''
 of $\, G_{14}^{8Dfcc}$,
the factor equivalent to $A_1$ in (\ref{last3}) to be an order-two 
self-adjoint operator with (\ref{RatSold8}) as the Wronskian. We 
should expect also the equivalent to $ \, E_7$ in (\ref{last3}) 
to be self-adjoint with even order greater than two.
It is tempting to find the ``canonical decomposition'' of  
$\, G_{14}^{8Dfcc}$, and see whether the order of the ``last'' factor 
(i.e. the equivalent of  $ \, E_7$ in (\ref{last3}))
is equal to the dimension $d= \, 8$ as we conjectured in~\cite{2015-LGF-fcc7}.

Indeed, the ``canonical decomposition''~\cite{2015-Canonical-Special} 
of  $\, G_{14}^{8Dfcc}$ is
\begin{eqnarray}
\label{decompG14}
\hspace{-0.95in} && \quad \,  \, 
G_{14}^{8Dfcc} \, = \,\, \, (A_2  \cdot B_2  \cdot C_2  \cdot    D_8 \,\,+ 
A_2  \cdot B_2\,\, + C_2  \cdot D_8\, \,+ A_2  \cdot D_8\, \, + 1) \cdot \, r(x), 
\end{eqnarray}
where $\,r(x)$ is a rational function, and where all the factors 
are {\em self-adjoint}, with the indices indicating the order.

The starting relation to obtain this decomposition is the homomorphism that
maps the solutions of $\, G_{14}^{8Dfcc}$ to the solutions of the adjoint:
\begin{eqnarray}
\label{homoG14d8}
\hspace{-0.95in}&& \quad   \quad \quad     \quad  \quad \quad 
adjoint(L_{12}) \cdot \, G_{14}^{8Dfcc} 
\, \,  \, = \, \, \,  \,   adjoint (G_{14}^{8Dfcc}) \cdot \, L_{12}  
\end{eqnarray}
The sequence of Euclidean right divisions (the indices indicate the orders)
\begin{eqnarray}
\label{euclid}
\hspace{-0.95in}    
G_{14}^{8Dfcc} \,  = \,  A_{2} \cdot \,L_{12}\, + \, L_{10}, \quad \,  
L_{12} \,  = \,  B_{2} \cdot \,L_{10}\, + \, L_{8},  \quad  \,  
L_{10} \,  = \, C_{2} \cdot \,L_{8}\, + \, r(x),   
\end{eqnarray}
and substitutions, give the decomposition (\ref{decompG14}).
We have shown in \cite{2015-Canonical-Special} that the factors 
$\,A_2$, $\, B_2$, $\,C_2$ are automatically self-adjoint. The 
sequence ends when the rest of the last Euclidean 
right division is a rational function: one, then, obtains the order-8
self-adjoint operator $ \,D_8 = \, L_8/r(x)$.
If the exterior square of $ \, G_{14}^{8Dfcc}$ was of the generic order 
$91$, and annihilated a rational function, the sequence of right 
divisions would continue, and the last Euclidean right division would be 
$\,\, L_4\,  =\,\,  F_2 \cdot L_2\, \, + r(x)$.

\subsection{The differential Galois group $G_{18}^{9Dfcc}$}
\label{G18}

Similar calculations performed on the operator $\, G_{18}^{9Dfcc}$ 
show that the symmetric square is of order $170$, instead of 
the generically expected order $\, 171$. 
The differential Galois group of the operator $\, G_{18}^{9Dfcc}$ 
is included in $\, SO(18, \,\mathbb{C})$.

The symmetric square of the adjoint of $\, G_{18}^{9Dfcc}$ 
annihilates the rational function
\begin{eqnarray}
\hspace{-0.75in} && \quad \quad \quad \, 
sol_R (sym^2 (adjoint (G_{18}^{9Dfcc}))) 
\,\,  =\, \, \, \,
{\frac{x^{28} \cdot \, P_{260}(x) \cdot \, S_9(x)^2}{ P_{18}(x)^2}}, 
\end{eqnarray}
where $ \, P_{18}(x)$ is the apparent polynomial of the operator
 $ \, G_{18}^{9Dfcc}$, $ \, S_9(x)$ is a degree-9
polynomial corresponding to the finite singularities given 
in \ref{sing891011},  and $ \, P_{260}$ is a degree-260 polynomial.

Heavy calculations give the canonical decomposition~\cite{2015-Canonical-Special}
of $ \, G_{18}^{9Dfcc}$ as (again, the indices denote the order):
\begin{eqnarray}
 \hspace{-0.95in} && \quad \quad \,\,
G_{18}^{9Dfcc} \, \, =\, \, \, 
(A_1 \cdot B_1 \cdot C_1 \cdot D_1 \cdot E_1 \cdot F_1 
  \cdot G_1 \cdot H_1 \cdot I_1 \cdot  J_9  \,\, + \cdots) \cdot \, r(x).
\end{eqnarray}
All the factors are self-adjoint.
The decomposition contains 89 terms and is obtained through a sequence 
of nine Euclidean right divisions starting from the homomorphism that 
maps the solutions of $ \, G_{18}^{9Dfcc}$ to the solutions of the adjoint:
\begin{eqnarray}
\label{adjG18}
\hspace{-0.95in}&& \quad   \quad \quad     \quad  \quad \quad 
adjoint(L_{17}) \cdot \, G_{18}^{9Dfcc} 
\, \,  \, = \, \, \,  \,   adjoint\left(G_{18}^{9Dfcc}\right) \cdot \, L_{17}. 
\end{eqnarray}
Here also, we see that the conjecture of~\cite{2015-LGF-fcc7} 
is verified. The order of the last self-adjoint factor (i.e. $\,J_9$) 
is equal to the dimension of the lattice, $\,d=\,9$.

\subsection{The differential Galois groups of $ \,G_{22}^{10Dfcc}$, $ \, G_{27}^{11Dfcc}$
and $ \, G_{32}^{12Dfcc}$}
\label{10Dfcc11Dfcc}

The detailed calculations done for the decompositions of  $\, G_{14}^{8Dfcc}$
and $ \, G_{18}^{9Dfcc}$ are too huge to be performed on 
$ \,G_{22}^{10Dfcc}$  and $ \, G_{27}^{11Dfcc}$.
However, it is straightforward to obtain that the exterior square of 
$ \, G_{22}^{10Dfcc}$ is of order $230$, 
instead of the generic order $231$.
The differential Galois group of the linear differential operator $ \, G_{22}^{10Dfcc}$ 
is included in $\, Sp(22, \,\mathbb{C})$.
The symmetric square of $ \, G_{27}^{11Dfcc}$ is of order $377$, 
instead of the generic order $378$.
The differential Galois group of the operator $G_{27}^{11Dfcc}$ 
is included in $\, SO(27, \,\mathbb{C})$.
Also, the exterior square of  the known modulo a prime $ \, G_{32}^{12Dfcc}$, 
is of order $495$ instead of the generic order $496$.
The differential Galois group of the operator $G_{32}^{12Dfcc}$ 
is included in $\, Sp(32, \,\mathbb{C})$.

\section{Coupling of lattices}
\label{sec:coupling}

In Section~\ref{computdetails}, we mentioned that some recurrences on the
coefficients $ \, t_d(n,p)$ have been used to improve the efficiency of
the computations.
Recall that to obtain the recursion relation giving the coefficients $ \, t_d(n,p)$, 
we introduced~\cite{2015-LGF-fcc7}
\begin{eqnarray}
\hspace{-0.95in}&& \quad   \quad    \, \,
\zeta_d \,\, = \, \,\,  \,  
\sum_{i=1}^d \, \sum_{j=i+1}^d \, \cos(k_i) \cdot \, \cos(k_j), 
\, \,\quad   \quad     \, \,\, 
\sigma_d \,\, = \, \, \,\,   \sum_{i=1}^d \, \cos(k_i), 
\end{eqnarray}
in terms of which the coefficients $ \, t_d(n,p)$ are given by
\begin{eqnarray}
\label{Tdfirst}
t_d \left(n, p \right) \, \,  \,  \,=\, \,\, \, \,  
 4^{n+p} \cdot \, \bigl\langle \zeta_d^n \cdot \sigma_d^{2 p} \bigr\rangle
\end{eqnarray}
where the symbol $\langle\cdot\rangle$ means that the integration 
on the variables $\,k_j$, occurring in the integrand, has been performed
(with the normalization $\pi^d$). 

It is straightforward to see that the coefficients $ \, t_d(n,p)$ 
correspond to the coefficients in the expansion around $ \, (0,0)$ 
of the $d$-dimensional integral 
\begin{eqnarray}
\label{integralTdzy} 
\hspace{-0.75in} && \quad \quad \quad \quad    
T_d (z, y) \,\,=\,\,\,\,
{\frac{1}{\pi^d}}\, \int_0^\pi \cdots \,
 \int_0^\pi \, {\frac{\mathrm{d}k_1\,  \cdots \,\,  \mathrm{d}k_d}{(1\, - 4 z \, \zeta_d) 
\cdot \, (1\,- 4 y  \, \sigma_d^2)}} 
\nonumber \\
 \hspace{-0.75in} && \quad \quad \quad \qquad \quad \quad   
\,\, = \,\,\,
\sum_{n=0} \, \sum_{p=0} \,\,  \, z^n \cdot \, y^p \cdot \, t_d(n,p), 
\end{eqnarray}
which gives the $ \, LGF_d(z)$ of the $d$-dimensional fcc lattice 
for $ \, y= \, 0$, and the $ \, LGF_d(y)$ of the $d$-dimensional simple 
cubic lattice for $ \, z= \, 0$. Note that for the simple 
cubic lattice, $ \, \sigma_d^2$  should be $ \, \sigma_d$. The expansion 
of the LGF of the simple cubic lattice corresponds to the expansion 
of (\ref{integralTdzy}) with $ \, z= \, 0$
and $ \, y = \,  (x/d)^2/4$, where $\, x$ is the expansion parameter.

\vskip 0.2cm

From the computation of $ \, t_d(n,p)$ for some values of $\, d$, we infer the
first terms of the expansion around $ \, (0,\, 0)$ of $ \, T_d(z,y)$ 
\begin{eqnarray}
\hspace{-0.95in} && 
T_d (z,y) \, \,=\,\,\, 
1\, \,\,   +2\,  d \cdot \,y\,\,   +2\,  d \cdot \, (d-1) \cdot \, {z}^{2}\,\, 
+4\, d \cdot \,(d-1) \cdot \, zy 
\, \,  +6\,  d \cdot  \, (2\, d -1) \cdot \,  {y}^{2}
 \nonumber \\
\hspace{-0.95in} &&  \quad \quad
+8\,  d \cdot \, (d-1)  \, (d-2)\cdot \,  {z}^{3} \, \, 
+4\,  d \cdot \, (d-1)  \, (5\, d -7)\cdot \, {z}^{2} y \, \, \, 
+ 48\,  d \cdot \, (d-1)^{2} \cdot \, z \, {y}^{2}
 \nonumber \\
\hspace{-0.95in} &&  \quad \quad
+20\,  d \cdot \, (4+6\,{d}^{2} -9\,  d) \cdot \, {y}^{3} 
\, \, \,  +  \, \cdots
\end{eqnarray}

\vskip 0.2cm

Even if there is no obvious lattice corresponding to the Green function 
(\ref{integralTdzy}), we found that it might be worthy to analyze the bivariate
series $ T_d (z,y)$, {\em per se}.

In the sequel, we address \footnote[1]{See e.g.~\cite{Evans-2010} for an 
introduction on partial differential equations.}
 the {\em system of linear partial differential 
equations} (PDE) that annihilates the $ \, D$-finite bivariate series 
$ \, T_d(z,y) \, $ for $ \, d= \, 2$.

\section{Partial differential equations for $\, T_2(z,y)$}
\label{partialdiff}

To find the PDEs that annihilate $ \, T_2(z,y)$ we can either proceed as for the linear 
ODEs, i.e. by the guessing method, or apply the creative telescoping
technique~\cite{zeilberger-ct,koutschan-2010,koutschan-survey};
the latter is computationally more costly, but provides a certificate of correctness
of the obtained differential equations. For example~\cite{koutschan-2013}, it 
was powerful enough to find {\em and prove} the ODEs satisfied by the LGF for 
$d= \, 4, \, 5, \, 6$, but failed for $ d \, \geq \, 7$.
All the differential equations mentioned in this and the following 
sections are also available in electronic form~\cite{koutschan-webpage}.

In order to apply the guessing method
we assume a partial differential equation 
of order $ \, Q$ in the homogeneous partial derivatives 
$\, z \cdot \partial/\partial z$ 
and $\, y \cdot \partial/\partial y$, with polynomials in $ \, z$
 and $\,y$ of degree $\,D$ that annihilates $ \, T_2(z,y)$:
\begin{eqnarray}
\hspace{-0.95in}&& \quad   \quad    
\sum_{q=0}^Q \, \sum_{n=0}^D \, \sum_{p=0}^D \, \,\, 
 a_{n,p}^{(q)} \cdot \, z^n \cdot \, y^p \cdot \, 
 \Bigl(z \cdot {\partial \over \partial z}\Bigr)^q \cdot 
\, \Bigl(y \cdot {\partial \over \partial y}\Bigr)^{Q-q} \cdot T_2 (z,y) 
\,\,  \,= \,\,\,\,  0. 
\end{eqnarray}
This linear system fixes the coefficients $ \, a_{n,p}^{(q)}$, and leaves 
some of them free. The number of non-fixed coefficients is the 
number of PDEs with order $ \, Q$ and degree $ \, D$.
If all the coefficients are such that $ \,a_{n,p}^{(q)} = \, 0$, we increase 
the order $ \, Q$ and/or the degree~$\,D$.

For $Q= \,1$, and various increasing values of $ \,D$, all the coefficients
are such that  $ \,a_{n,p}^{(q)} = \,0$.
For $Q= \,2$ and $D= \,2$, there is only one PDE, that we denote $ \,PDE_2$.
For $Q= \,3$ and $D= \,1$, there are only two PDEs, called 
$ \,PDE_3^{(1)}$ and $ \,PDE_3^{(2)}$. Note that 
{\em there is no concept of ``minimal order'' for PDEs}, while 
there is one for ODEs.
Instead, one can consider a {\em Gr\"obner basis} (see~\ref{groebner}) in the ring of partial
differential operators, as is demonstrated in Section~\ref{T2GB}.
It is obvious that there are as many PDEs that annihilate $ \,T_2(z, y)$ 
as we wish (namely, all elements of the left ideal~$\ann(T_2)$,
see \ref{groebner}).
For instance, for $Q= \,4$ and $D= \,1$, we obtain 
five PDEs, three of them are of order four and two of them are 
combinations of $ \,PDE_3^{(1)}$ and $ \,PDE_3^{(2)}$.

\subsection{Two PDEs for $ \, T_2(z,y)$}
\label{TwoPDE}

With the notation
\begin{eqnarray}
\label{mixed}
\hspace{-0.75in}&& \qquad  \quad  \qquad   \quad   \quad  \quad  
D_{zy}^{(n,p)} \, \, =\,\, \,\,
  {\frac{\partial^{n+p}}{\partial z^n \, \partial y^p}}, 
\end{eqnarray}
the system of two partial differential operators for  $ \, T_2(z,y)$ reads
\begin{eqnarray}
\label{firstPDEforT2}
\hspace{-0.95in} && \, \, \, 
PDE_3^{(1)} \, \, =\, \,  \, \,  
2 \,  y^3 \cdot \, (16 y-1)\cdot \,  D_{zy}^{(0,3)} \, \,\,  
+3 y^2 z \cdot \, (16 z y+12 y-1)\cdot \,  D_{zy}^{(1,2)} \,  
 \nonumber \\
 \hspace{-0.95in} && \quad \quad 
+y z^2 \cdot \, (48 z y+12 y-1-2 z)  \cdot \,  D_{zy}^{(2,1)} \, \, 
+12 y z^2 \cdot \, (4 z+1)\cdot \,  D_{zy}^{(2,0)} \, 
 \nonumber \\
\hspace{-0.95in} && \quad \quad 
+4 y z \cdot \, (60 z y+24 y-1-z)\cdot \,  D_{zy}^{(1,1)} \, \, 
+2 y^2\cdot \,  (24 z y+80 y-3)\cdot \,  D_{zy}^{(0,2)}
 \nonumber \\
\hspace{-0.95in} && \quad \quad 
+24 y z \cdot \, (6 z+1)\cdot \,   D_{zy}^{(1,0)} \, \, 
+2 y \cdot \,  (68 y-1+72 z y)\cdot \,  D_{zy}^{(0,1)} \, \, 
 \nonumber \\
\hspace{-0.95in} && \quad \quad 
+8 y \cdot \, (6 z+1) \cdot \, D_{zy}^{(0,0)}, 
\end{eqnarray}
and: 
\begin{eqnarray}
\label{secondPDEforT2}
\hspace{-0.95in} &&  
PDE_3^{(2)} \, =\,  \, \, \,   
2 \,  y^3 \cdot \, (16 y-1) \cdot \,  D_{zy}^{(0,3)} \, \, 
+y^2 z \cdot \, (24 y+32 z y-3+4 z)\cdot \,   D_{zy}^{(1,2)} \, 
\nonumber \\
\hspace{-0.95in} &&  \quad \quad 
+y z^2 \cdot \, (32 z y-1+8 y)\cdot \,   D_{zy}^{(2,1)} \, \, 
+8\,  y z^2 \cdot \, (4 z+1) \cdot \,  D_{zy}^{(2,0)} \,
\nonumber \\
\hspace{-0.95in} &&  \quad \quad 
+4\,  y z \cdot \, (40 z y+16 y-1+z) \cdot \,  D_{zy}^{(1,1)} \, \, 
+2\,  y^2 \cdot \, (16 z y+80 y-3+2 z)\cdot \,   D_{zy}^{(0,2)}
  \nonumber \\
\hspace{-0.95in} && \quad \quad 
+16 \, y z \cdot \, (6 z+1) \cdot \,  D_{zy}^{(1,0)} \, \, 
+2\,  y\cdot \,  (68 y+2 z+48 z y-1) \cdot \,   D_{zy}^{(0,1)} \, \, 
 \nonumber \\
\hspace{-0.95in} && \quad \quad 
+8\,  y \cdot \, (4 z+1) \cdot \,  D_{zy}^{(0,0)}.
\end{eqnarray}

These two operators annihilate (by construction) a finite truncation of the power series $\, T_2(z,y)$,
where the truncation index has been chosen such that it is very likely that they also annihilate the
infinite series $\,T_2(z,y)$. That they indeed annihilate $\,T_2(z,y)$ will be made rigorous in
Section~\ref{T2GB}. In particular, it follows that $ \,PDE_3^{(1)}$ and $ \,PDE_3^{(2)}$ are compatible.
By assuming a common solution of the form 
\begin{eqnarray}
\label{bivariateform}
\hspace{-0.95in}&& \qquad  \qquad   \qquad  \, \, 
\sum_{n=0} \sum_{p=0}\,\, \,  b_{n,p} \cdot \, z^n \cdot \, y^p, 
\end{eqnarray}
we find a unique solution to the system of PDEs that 
identifies with $ \, T_2(z,y)$, up to
an overall constant. Only one constant means that if we switch 
to recurrences on the coefficients we will need only one 
initial condition.

The recursions for $\, t_2(n,p)= \, U(n,p) \, $ are
\begin{eqnarray}
\hspace{-0.85in} &&  \quad   
-4 \cdot \, (8 p^2+17 p+9 n p+9 n+8+3 n^2) \cdot \, U(n+1,p) \,
 \nonumber \\
\hspace{-0.85in} && \quad  \quad  \quad   
 +2 \cdot \, n (n+1) \cdot \, U(n,p+1)\,\,
 +(n+p+2) (2 p+n+3) \cdot \,  U(n+1,p+1) \, 
\nonumber \\
\hspace{-0.85in} && \quad  \quad  \quad   
 -48  \cdot \, (n+1) (n+p+1) \cdot \,  U(n,p)
\, \, \,=\,\,\, 0,  
\end{eqnarray}
and
\begin{eqnarray}
\hspace{-0.95in} && 
8 \cdot \, (3+4 p^2+7 p+3 n p+3 n+n^2)\cdot \, U(n+1,p) \, 
\nonumber \\
\hspace{-0.85in} && \quad  \quad    
+4 \cdot \, (n+1) (p+1)\cdot \, U(n,p+1) 
\, \, +(p -n+2) (2 p+n+3)\cdot \, U(n+1,p+1) \, 
\nonumber \\ 
\hspace{-0.95in} && \quad  \quad    
+32 \cdot \, (n+1) (n+p+1)\cdot \, U(n,p) 
\,\,  \,=\,\,\,0, 
\end{eqnarray}
with the auxiliary recurrence:
\begin{eqnarray}
\hspace{-0.75in} && \quad  \qquad  
(p+1)^2\cdot \,  U(0,p+1) \,\,\, 
 -4 \cdot \, (2 p+1)^2\cdot \,  U(0,p)
 \, \,\,= \,\,\, \, 0. 
\end{eqnarray}
With the coefficient $ \, U(0,0)$ given, these three recurrences 
generate all the $ \, U(n,p)$. 

\subsection{One PDE for $\, T_2(z,y)$}
\label{OnePDE}
 
There is only one {\em partial differential operator}
 of order $Q= \, 2$ and degree $\, D= \, 2$ 
that annihilates $ \, T_2(z,y)$
\begin{eqnarray}
\label{onlyPDEforT2}
\hspace{-0.95in} && 
PDE_2 \,\, =\,\, \,\, 
z^2 \cdot \, (4z-1)(4z+1)(4y-1) \cdot \, D_{zy}^{(2,0)}
 \nonumber \\
\hspace{-0.95in} && \quad  \quad 
+ zy \cdot \, (64z^2y-16z^2-4z-20y+3) \cdot \, D_{zy}^{(1,1)}
 \nonumber \\
\hspace{-0.95in} &&\quad  \quad 
-2  \, y^2 \cdot \,  (16y-1) \cdot \,  D_{zy}^{(0,2)} \, \, 
+ z \, (192z^2y-48z^2-32zy-12y+1) \cdot \,  D_{zy}^{(1,0)} 
\nonumber \\
\hspace{-0.79in} && \quad  \quad 
+2 \, y \cdot \, (32z^2y-8z^2-32y-24zy-2z+1)  \cdot \,  D_{zy}^{(0,1)} 
\nonumber \\
\hspace{-0.95in} && \quad  \quad 
- (16z^2+32zy+8y-64z^2y)  \cdot \, D_{zy}^{(0,0)}, 
\end{eqnarray}
which acting on the bivariate form (\ref{bivariateform}) generates, 
remarkably, a unique solution that identifies with $ \, T_2(z,y)$, 
up to an overall constant.

The coefficients $ \,\,  t_2(n, p)= \, U(n, p)\, $ are given 
by the recursion
\begin{eqnarray}
\hspace{-0.95in} && \quad  \quad 
-(20 n p+72 p+40+24 n+32 p^2+4 n^2) \cdot \,  U(n+2,p)
 \nonumber \\
\hspace{-0.95in} && \quad  \quad 
+64 \, (n+1) (n+p+1) \cdot \,  U(n,p)\, \,\,  
-4 (p+1) (n+2) \cdot \,  U(n+1,p+1)
 \nonumber \\
\hspace{-0.95in} && \quad  \quad 
+(p+3+n) (2 p+4+n) \cdot \,  U(n+2,p+1)
 \\
\hspace{-0.95in} && \quad \quad 
-(64+48 p+32 n) \cdot \,  U(n+1,p)\, \,\,  
-16 \cdot \, (n+1) (n+2+p) \cdot \,  U(n,p+1)
 \, \, = \, \, \, 0, \nonumber 
\end{eqnarray}
with the auxiliary recursions:
\begin{eqnarray}
\hspace{-0.95in} && \quad  \,  
(n+2)^2  \cdot \, U(n+2,0) \, \,  -16  \cdot \, (n+1)^2  \cdot \, U(n,0) 
\, \, = \, \, \, 0,
 \quad \quad \quad  \,   U(1, 0) \, = \, \,  0, 
  \nonumber \\
\hspace{-0.95in} && \quad  \,  
(p+1)^2 \cdot \,  U(0,p+1) \, \,\,  -4 \, (2 p+1)^2 \cdot \,  U(0,p)
 \,  \, = \, \,\,  0,
 \\
\hspace{-0.95in} && \quad  \,  
(p+2) (2 p+3) \cdot \,  U(1,p+1) \,\,  \, 
-4  \, (p+1) (5+8 p) \cdot \, U(1,p)
\nonumber \\
\hspace{-0.95in} && \quad   \quad  \qquad  \qquad   \qquad  \qquad
 = \, \, \, \,  16 \cdot \,  (2+3 p)  \cdot \,  U(0,p)
 \, \,  \,   +4  \, (p+1) \cdot \,  U(0,p+1).
 \nonumber 
\end{eqnarray}
Here also, these recurrences generate all the coefficients starting 
with $ \, U(0,0)$.
 
\subsection{On the logarithmic solutions}
\label{log}

The system of PDEs given in (\ref{firstPDEforT2}, \ref{secondPDEforT2})
has no logarithmic solution of the form
\begin{eqnarray}
\label{logForm}
\hspace{-0.75in} && \quad \qquad \quad
\sum_{n=0} \sum_{p=0}^n \,\, F_{n,p}(z,y) \cdot \, \ln(z)^n \cdot \, \ln(y)^{n-p}, 
\end{eqnarray}
where $ \, F_{n,p}(z,y)$ are analytic at $(0,0)$ bivariate series.

In contrast, the PDE given in (\ref{onlyPDEforT2}) {\em seems to have no bound in the
summation on} $ \, n$ (we obtained logarithmic solutions up to $ \, n=\, 17$).
Furthermore, we find that the logarithms $\, \ln(z)$, and $\, \ln(y)$, appear
in the solutions as:
\begin{eqnarray}
\label{logFormmu}
\hspace{-0.75in} && \quad \qquad \quad
\sum_{n=0} \sum_{p=0}^n \, \, 
F_{n,p}(z,y) \cdot  (\ln(z)\, - \mu \cdot \ln(y))^{p}.
\end{eqnarray}

The number of logarithmic solutions depends now on the value of $\mu$.
For $\,\mu = \, 1$ and $\,\mu= \, 1/2$, we find no bound to $ \, n$ 
(in our calculations, we reached $\,n=\,17$).
For any other value of $\,\mu\, \ne \,1, \,1/2$, there is only one logarithmic 
solution i.e. $\,n=\,1$. For generic $\mu$ the solutions are $\,T_2(z, y)$ 
and\footnote{Note that the derivative with respect to $\mu$ of the logarithmic
solution is also a solution.}:
\begin{eqnarray}
\hspace{-0.95in} && \quad 
T_2(z,y) \cdot (\ln(z) \,  - \mu \cdot \ln(y))\,\,
\nonumber \\
\hspace{-0.95in} && \quad \quad  
 + \Bigl( {1 \over 2}\,\, +2 \,\mu \,\,\, -4\,\mu \cdot \,z\,\,
 +2\,\cdot (3+5\,\mu) \cdot  {z}^{2} \,\,\, +4\,\cdot  (1-\mu) \cdot \,zy 
\nonumber \\
\hspace{-0.95in} && \quad \quad 
- (13+12\,\mu) \cdot  {y}^{2} \, \, \, -{\frac {304}{9}}\,\mu \cdot {z}^{3}\,  \,
+{2 \over 3} \cdot (33-16\,\mu) \cdot  {z}^{2}y \,\, \, +{\frac {8}{15}}
\cdot (31-90\,\mu) \cdot z {y}^{2} 
\nonumber \\
\hspace{-0.95in} && \quad \quad 
-{4 \over 9} \cdot (559+420\,\mu) \cdot  {y}^{3} \,\, \,  +\,  \cdots \Bigr).
\end{eqnarray}

\vskip 0.1cm

Let us address the details of the computations on how the values 
$\mu= \, 1$, and $\mu= \, 1/2$ appear.
Acting by $ \, PDE_2$ on the form (\ref{logFormmu}) rewritten as
\begin{eqnarray}
\label{logFormmuREW} 
\hspace{-0.65in} && \quad \quad \quad \quad
\,F_{n,n}(z,y) \cdot \, (\ln(z)\,  - \mu \cdot \, \ln(y))^{n}
 \, \,\, + \, \cdots, 
\end{eqnarray}
gives the choice of zeroing 
\begin{eqnarray}
\label{choice}
\hspace{-0.65in} && \quad \quad \quad \quad
 a_{0,0} \, \cdot \, (\mu -1/2) \, (\mu-1) \, \,=\,\, \, 0, 
\end{eqnarray}
where $\, a_{0,0}$ is the leading coefficient of the bivariate 
series $\, F_{n,n}(z,y)$. 
The choice $\mu=1$ (or $\mu=\, 1/2$) allows $\, a_{0,0}\, \ne\, 0$, which
permits $\,n$ to be higher. 
The choice $\,a_{0,0}=\,0$ will decrease the degree $n$, and the
process continues with $\,n-1$.

The choice (\ref{choice}) comes from the action of $\,PDE_2$ on 
(\ref{logFormmuREW}), and the leading coefficient of the expansion
to be cancelled is:
\begin{eqnarray}
\hspace{-0.95in}&&    \quad 
n \cdot \,(n-1) \cdot \, a_{0,0} \cdot \, (\mu -1/2) \, (\mu-1)\, 
\cdot \, (\ln(z) - \mu \cdot \, \ln(y))^{n-2} \, \, \,  + \,  O(z^1, y^1).
\end{eqnarray}

\subsection{Gr\"obner basis of PDEs for $\,T_2(z,y)$}
\label{T2GB}

A Gr\"obner basis for $\ann(T_2)$, the annihilating ideal of PDEs for $\,T_2(z,y)$, can
be obtained by applying Buchberger's algorithm to the input
$\bigl\{PDE_3^{(1)},\,PDE_3^{(2)}\,PDE_2\bigr\}$ (some basics about Gr\"obner
bases are given in~\ref{groebner}). Alternatively, we can compute the
annihilating ideal {\em from scratch}, i.e. from the integral
representation~(\ref{integralTdzy}) of $\,T_2(z,y)$, by the method of
creative telescoping. Both tasks can be performed with the HolonomicFunctions
package~\cite{koutschan-package} and yield the same result. The second approach,
however, gives an independent proof that the guessed PDEs presented in the
previous sections are correct.

The Gr\"obner basis of $\, \ann(T_2)$ (with respect to degree-lexicographic order
and $D_y\prec D_z$) consists of $3$ operators, whose supports are given as follows:
\begin{eqnarray*}
&& \bigl\{ D_{zy}^{(2,0)}, D_{zy}^{(1,1)}, D_{zy}^{(0,2)}, D_{zy}^{(1,0)}, D_{zy}^{(0,1)}, D_{zy}^{(0,0)} \bigr\}, \\
&& \bigl\{ D_{zy}^{(0,3)}, D_{zy}^{(1,1)}, D_{zy}^{(0,2)}, D_{zy}^{(1,0)}, D_{zy}^{(0,1)}, D_{zy}^{(0,0)} \bigr\}, \\
&& \bigl\{ D_{zy}^{(1,2)}, D_{zy}^{(1,1)}, D_{zy}^{(0,2)}, D_{zy}^{(1,0)}, D_{zy}^{(0,1)}, D_{zy}^{(0,0)} \bigr\}.
\end{eqnarray*}
Note that the first basis element is exactly $PDE_2$.
By investigating the {\em leading monomials} $D_{zy}^{(2,0)},D_{zy}^{(0,3)},D_{zy}^{(1,2)}$ one
immediately finds that there are $5$ monomials under the stairs, namely the monomials
$D_{zy}^{(1,1)}, D_{zy}^{(0,2)}, D_{zy}^{(1,0)}, D_{zy}^{(0,1)}, D_{zy}^{(0,0)}$, which cannot be
reduced by either of the leading monomials. We say that $\ann(T_2)$ has holonomic rank~$5$.
Hence one could expect that $5$ initial conditions have to be given to identify the
particular solution $\,T_2(z,y)$. As discussed before, we remarkably need only one initial
condition.

\section{Ordinary differential equations for $\, T_2(z, y)$}
\label{ODET2}

The {\em bivariate series} $\, T_2(z, y)$ may be seen as depending 
on the variable $ \,z$ (or $y$) where $y$ (or $z$) is a parameter. By 
derivation of the PDE system, and elimination of the unwanted derivatives, 
one obtains a linear ODE on the variable $z$ (or $y$) that annihilates $ \, T_2(z, y)$.
Such elimination can be conveniently performed by using the Gr\"obner basis
presented in Section~\ref{T2GB}.

\subsection{ODE with the derivative on $\, z$ for $\, T_2(z,y)$}
\label{ODEzT2}

The linear ODE with the variable $ \, z$, that annihilates $ \, T_2(z,y)$,
 is of order five, and we call the corresponding operator $ \, L_5^{(z)}$,
(with the derivative $\, D_z = \, {\partial \over \partial z}$):
\begin{eqnarray}
\label{defL5z}
\hspace{-0.95in}&& \quad \qquad   \qquad
L_5^{(z)}\,\,   = \, \, \, \sum_{n=0}^{5} \, P_n(z, y) \cdot \, D_z^n.
\end{eqnarray}

The polynomial in front of the highest derivative is
\begin{eqnarray}
\hspace{-0.95in}&& \quad   \quad   \quad 
{z}^{2} \cdot \, (4\,z-1) \cdot  \, (4\,z+1) \cdot  \, (z-4\,y) \cdot 
 \, (16\,{z}^{2}y+y+8\,z y-4\,{z}^{2}) \cdot \, P_{app}, 
\end{eqnarray}
where $\, P_{app}$ carries apparent singularities:
\begin{eqnarray}
\hspace{-0.95in}&&    \quad 
P_{app} \, \,=\, \, \,  \,  \, 
192\,y \cdot \, (4\,y-1) \cdot \,  {z}^{5} \, \,  \, 
- \, (128\,{y}^{2} +32\,y-12) \cdot \, {z}^{4} \,  \, \, 
+4\,y \, (80\,y-19) \cdot \, {z}^{3}
 \nonumber \\
\hspace{-0.95in}&& \quad   \quad  \quad  \quad   \quad 
- (40\,{y}^{2}+4\,y-1) \cdot \,  {z}^{2} \,\, \, 
+y \cdot \, (y-1)  \cdot \,  z \, \,\, \,  +{y}^{2}.
\end{eqnarray}

The factorization of the order-five linear differential 
operator $ \, L_5^{(z)}$ reads 
(the indices are the orders)
\begin{eqnarray}
\label{directsumdecompL}
\hspace{-0.75in}&&    \quad  \qquad  \,  \,  \, 
L_5^{(z)}\, \,=\, \, \, \,
\left( L_1^{(2)} \cdot L_1^{(1)} \right) \oplus \left( L_1^{(3)} \cdot L_1^{(1)} \right)
 \oplus \left( L_2 \cdot L_1^{(1)} \right), 
\end{eqnarray}
where the four factors are given in \ref{factorizL5zN5y}.

\vskip 0.1cm

The solution of $ \, L_1^{(1)}$ reads:
\begin{eqnarray}
\hspace{-0.75in}&& \quad   \quad  \quad 
sol \left( L_1^{(1)} \right) \,\,=\,\,\,\,\,
\sqrt{ {\frac{z}{(z-4 y) \cdot \,
 (4 \cdot \, (4 y-1) \cdot  \, z^2 \, +8 z y \,+y) }} }. 
\end{eqnarray}
The second solution of $\, L_1^{(2)} \cdot L_1^{(1)}$ reads: 
\begin{eqnarray}
\hspace{-0.75in}&& \quad   \quad  \quad 
sol (L_1^{(1)}) \cdot \int \, 
{\frac{z^{-3/2}}{\sqrt{ (z-4 y) \, (4 \cdot \, (4 y-1) \cdot \,  z^2 \, +8 z y\, +y) }} } 
\cdot \, \mathrm{d}z. 
\end{eqnarray}
The integral can be evaluated in terms of the incomplete elliptic integrals,
so that the second solution of $ \, L_1^{(2)} \cdot \, L_1^{(1)}$ reads 
\begin{eqnarray}
\hspace{-0.95in}&& 
sol (L_1^{(1)}) \cdot \, 
\Bigl({\frac { (4\,y-\sqrt {y}) }{2 {y}^{2}}} \cdot \,E(z_1, z_2)
-{\frac { 4 \cdot \, (8\,y-1-2\,\sqrt {y}) }{y \cdot \, (16\,y-1) }}
\cdot  \, F(z_1, z_2)\Bigr) 
\, \,  \,   \, 
-{\frac {2 }{ (z-4\,y)\, y}}, 
\nonumber 
\end{eqnarray}
with
\begin{eqnarray}
\hspace{-0.95in}&& \qquad \quad \quad  \,   \, 
z_1 \,=\, \, 
 \sqrt {{\frac { (4\,\sqrt {y}-1)^{2} z}{z-4\,y}}}, 
\qquad \, \, \, 
z_2 \,=\, \, 
\sqrt {{\frac { (4\,\sqrt {y}+1)^{2}}{ \left( 4 \,\sqrt {y}-1 \right) ^{2}}}} 
\end{eqnarray}
and where $\, E$ and $\, F$ are the incomplete elliptic integrals:
\begin{eqnarray}
 \hspace{-0.95in}&&  \, 
E( z, k)  \, \,=\, \,  \,
\int_0^{z} \, {\frac{\sqrt{1-k^2 t^2}}{\sqrt{1-t^2}}} \cdot \, \mathrm{d}t, 
\quad \, \,  \, 
F( z, k)  \, \,=\,\,   \,
\int_0^{z} \, {\frac{1}{ \sqrt{1-k^2 t^2} \, \,  \sqrt{1-t^2}}} \cdot \,\mathrm{d}t.
\end{eqnarray}

\vskip 0.1cm

The second solution of $ \,\, L_1^{(3)} \cdot L_1^{(1)} \, $ can be written 
as the general {\em Heun function}
\begin{eqnarray}
\hspace{-0.75in}&& \quad \quad \qquad\,  \,
f(z) \cdot \, 
Heun \Bigl(a, \, q, \, {{1} \over {2}},\, 
1,\, {{3} \over {2}}, \, {{1} \over {2}}, \,  g(z)\Bigr), 
\end{eqnarray}
with
\begin{eqnarray}
\hspace{-0.75in}&& \quad \quad \quad \,  \,
f(z)\,  \,=\, \, \, 
{\frac {z\cdot  \, \sqrt {4\,z \, \sqrt {y} \, \, +\sqrt {y}\,-2\,z}}{ 
(16\,{z}^{2}y \,+y +8\,z y\,-4\,{z}^{2}) \cdot \, \sqrt {z-4\,y}}},
 \nonumber \\
\hspace{-0.75in}&& \quad \quad \quad \,  \,
g(z) \, \,= \, \, \, 
{\frac { z y \cdot \, (64\,z y+16\,y-12\,z+1) \, \, 
 -2\, z \cdot \, \sqrt {y} \cdot \, (z-4\,y) }
{ 4\, y \cdot  \, (16\,{z}^{2}y+y+8\,z y-4\,{z}^{2})  }}, 
\end{eqnarray}
and:
\begin{eqnarray}
\hspace{-0.75in}&& \quad \quad \quad \quad \quad \,  \,
a  \,\,  =  \, \, \,  {1 \over 2} \,\,  +{\frac {16\,y+1}{16 \sqrt {y}}},
 \qquad \,  \, 
q \, \,  =  \, \,  \, {\frac {1+a}{4}}. 
\end{eqnarray}

\vskip 0.1cm

The solution of $ \, \, L_2 \cdot L_1^{(1)} \, $ which is not solution
of $ \,L_1^{(1)} \, $ can be written as 
\begin{eqnarray}
\hspace{-0.75in}&& \quad \quad \quad \quad \quad \, \, 
sol (L_1^{(1)}) \cdot \int \, 
{\frac{ sol  L_2)}{sol (L_1^{(1)}) }} \cdot \, \mathrm{d}z, 
\end{eqnarray}
where one of the solutions of $L_2$ reads
\begin{eqnarray}
\hspace{-0.95in}&&   
sol (L_2) \,\,  =\, \, \, \, 
{\frac {z \cdot \, (16\,z^2-1)  \, 
( 64\,{z}^{3}y-80\,{z}^{2}y+16\,{z}^{2}-36\,zy-3\,y) }{
3 y \, (z -4\,y)  \, (y+8\,zy+16\,{z}^{2}y-4\,{z}^{2})}} 
\cdot \, {\frac{d H(z)}{dz}}
 \nonumber \\
\hspace{-0.95in}&& \quad \, \,  
+\, 4 \, \cdot \,{\frac { z \cdot \, 
(256\,{z}^{4}y +112\,{z}^{3} -512\,{z}^{3}y
 -224\,{z}^{2}y +z-32\,zy-3\,y) }{ y \cdot \, (z -4\,y) 
 \, (y+8\,zy+16\,{z}^{2}y-4\,{z}^{2})}} \cdot  \, H(z), 
\end{eqnarray}
where $\, H(z)$ is the hypergeometric function
\begin{eqnarray}
\hspace{-0.75in}&& \quad \quad \quad \quad \quad \quad \quad 
H(z)\, \,   \, = \,\, \,  \, 
 {_2}F_1 \Bigl( [{{3} \over {2}}, \, {{5} \over {2}}],  \, [1], \,  16  \, z^2 \Bigr). 
\end{eqnarray}

\subsection{ODE  with the derivative on $\, y$ for $\, T_2(z,y)$}
\label{ODEyT2}

The {\em bivariate series} $ \, T_2(z,y)$, where $ \, z$ is a parameter, 
is annihilated by an order-five linear differential operator
$ \, N_5^{(y)}$ (with the derivative $\, D_y = \, {\partial \over \partial y}$):
\begin{eqnarray}
\label{defN5y}
\hspace{-0.95in}&& \quad \qquad   \qquad
N_5^{(y)}\,\,   = \, \, \, \sum_{n=0}^{5} \, Q_n(z, y) \cdot \, D_y^n.
\end{eqnarray}

The polynomial, in front of the highest derivative, reads
\begin{eqnarray}
\hspace{-0.95in}&& \quad \quad  \quad \quad  
{y}^{2} \cdot \, (16\,y-1) \cdot \, (z-4\,y)  
\cdot \, (16\,{z}^{2}y+y+8\,z y-4\,{z}^{2}) \cdot \, P_{app}, 
\end{eqnarray}
where $\, P_{app}$ carries {\em apparent singularities}:
\begin{eqnarray}
 \hspace{-0.95in}&& \quad \quad  \quad  
P_{app} \, \,=\, \, \, 
-12 \cdot \, (4\,z+1)  \cdot \, (32\,{z}^{2}-12\,z-1) \cdot \,  {y}^{3}
\nonumber \\
\hspace{-0.95in}&& \quad \quad    \qquad \quad  
\, + \, (1344\,{z}^{3}-8+20\,{z}^{2}-114\,z+64\,{z}^{4}) \cdot \, {y}^{2} 
 \\  
\hspace{-0.95in}&& \quad \quad    \qquad \quad  
- \, (1+12\,z-20\,{z}^{2} -264\,{z}^{3}+48\,{z}^{4}) \cdot \,  y\, \, \,
+ \, (32\,{z}^{2}-1-6\,z) \cdot \, z^2.
 \nonumber
\end{eqnarray}

The order-five operator $\, N_5^{(y)}$ has the following direct sum factorization
\begin{eqnarray}
\label{N5blabla} 
\hspace{-0.95in}&& \quad \quad  \quad \quad  \quad   
N_5^{(y)} \,\, =\,\,\,\,
 \left( N_1^{(2)} \cdot N_1^{(1)} \right) \oplus \left( N_1^{(3)} \cdot N_1^{(1)} \right) 
\oplus \left( N_2 \cdot N_1^{(1)} \right), 
\end{eqnarray}
where the four factors are given in \ref{factorizL5zN5y}. 

\vskip 0.1cm

The solution of $\,N_1^{(1)}$ reads:
\begin{eqnarray}
\hspace{-0.95in}&& \quad \quad    \qquad
sol (N_1^{(1)}) \, \,=\,\,\,\,
\sqrt{ {\frac{y}{(z-4 y) \, (4  \cdot \, (4 y-1) \cdot \, z^2 \, +8 z y \, +y) }} }.
\end{eqnarray}

The second solution of $ \, N_1^{(2)} \cdot N_1^{(1)}\, $ reads:
\begin{eqnarray}
\hspace{-0.95in}&& \quad \quad    \qquad
sol \, (N_1^{(1)}) \cdot \, \int \, 
{\frac{y^{-3/2} \cdot \,(16 y+1)}{\sqrt{ (z-4 y)  \cdot \,
 (4 \cdot \, (4 y-1) z^2 \, +8 z y +y) }} } \cdot \, \mathrm{d}y. 
\end{eqnarray}
The integral can be evaluated in terms of the incomplete elliptic integrals and 
the second solution of $ \,\, N_1^{(2)} \cdot N_1^{(1)} \,\, $ reads
\begin{eqnarray}
 \hspace{-0.95in}&& \, \, \,\,
sol \, (N_1^{(1)}) \cdot \, \Bigl( {\frac { (4\,z-1)}{2  \, \,{z}^{5/2}}} \cdot \, E(z_1,z_2)\,\,
+\, {\frac { (80\,{z}^{2}+1+8\,z) }{2  \,\, {z}^{5/2} \cdot \, (4\,z-1) }} 
\cdot \,F(z_1,z_2)\Bigr)
\, \, \,\, -{\frac {1}{2 {z}^{3}}}, 
\end{eqnarray}
with:
\begin{eqnarray}
 \hspace{-0.95in}&& \, \quad  \quad  \quad 
z_1 \,=\,\, \sqrt {-{\frac {16\,{z}^{2}y \, +y+8\,z y \,-4\,{z}^{2}}{4\, {z}^{2}}}},
\qquad  \, \, 
z_2\, =\,\, \sqrt {-{\frac {16 z}{ (4\,z-1)^{2}}}}. 
\end{eqnarray}

The second solution of $\,N_1^{(3)} \cdot N_1^{(1)}\, $ is a general Heun function
\begin{eqnarray}
\hspace{-0.75in}&& \, \quad  \quad  \qquad 
sol \, (N_1^{(1)}) \cdot \sqrt{y} \cdot \, 
 Heun \Bigl(a,\,   q,  \,  {{1} \over {2}},\,   1,  \, 
 {{3} \over {2}}, \,  {{1} \over {2}}, {\frac {4 y}{z}}\Bigr), 
\end{eqnarray}
with:
\begin{eqnarray}
\hspace{-0.75in}&& \, \quad  \quad  \quad \quad \quad  \, 
a \,  = \,  \, {\frac {16\,z}{(4 z+1)^2}}, \qquad \, \, \, 
q \,  =  \, \,  {\frac {1+a}{4}}.
\end{eqnarray}

\vskip 0.1cm

The solution of $ \, \, N_2 \cdot N_1^{(1)} \, $ which is not solution
of $ \,N_1^{(1)} \, $ can be written as 
\begin{eqnarray}
\hspace{-0.75in}&& \, \quad  \quad  \qquad \quad \quad 
sol \, (N_1^{(1)}) \cdot \int \, 
{\frac{ sol \, (N_2)}{sol \, (N_1^{(1)}) }} \cdot  \, \mathrm{d}y, 
\end{eqnarray}
where one of the solutions of $\, N_2$ reads
\begin{eqnarray}
\hspace{-0.95in}&&  \, 
sol \, (N_2)\,\,  = \, \, \, \,
{\frac {y \cdot  \, (16\,y-1) \cdot \, 
(64\,{z}^{2}y \,+48\,zy \,+16\,y \,-20\,{z}^{2} \,-3\,z) }{ 
3  \cdot \,{z}^{5} \cdot \, (z-4\,y)  \cdot \, 
(y+8\,zy+16\,{z}^{2}y-4\,{z}^{2}) }}  
\cdot \, {\frac{d H_y(y)}{dy}}
 \nonumber \\
\hspace{-0.95in}&& \, \quad  \quad \, \, \, \, 
+2 \, \cdot \,{\frac {256\,{z}^{2}{y}^{2} +384\,{y}^{2}z 
+112\,{y}^{2} -112\,{z}^{2}y-24\,zy+y-2\,{z}^{2}
}{  {z}^{5} \cdot \, (z-4\,y) 
 \cdot \, (y+8\,zy+16\,{z}^{2}y-4\,{z}^{2})}} \cdot \, H_y(y), 
\end{eqnarray}
where $\, H_y(y)$ is the hypergeometric function:
\begin{eqnarray}
\hspace{-0.75in}&& \, \quad  \quad  \quad  \quad  \quad 
H_y(y) \, \,=\, \,\,  \,
 {_2}F_1 \Bigl( [{{3} \over {2}}, \, {{5} \over {2}}], \,  [1], \,  16 \, y \Bigr).
\end{eqnarray}

\subsection{ The linear ODE on $\, z$ and $ \, y$ as PDE for $ \, T_2(z,y)$}
\label{ODEzyT2}

The linear differential equations corresponding to the operators $ \, L_5^{(z)}$ and
$ \, N_5^{(y)}$ act on $\,T_2(z,y)$ as a system of decoupled PDEs.
Both ODEs annihilate (as it should) the bivariate series $ \, T_2(z,y)$,
and they generate a unique common bivariate series solution, 
analytic at $ \, (0,0)$, that identifies with $ \, T_2(z,y)$.

As for the solutions of the three PDEs of the previous sections, one 
remarks that $ \, sol \, (L_1^{(1)})\, $ and $ \, sol (N_1^{(1)})\,$ 
have simple structures, and we can check that
\begin{eqnarray}
\label{solPDET2}
\hspace{-0.75in}&& \, \quad  \quad \quad \quad  \quad 
\sqrt{ {\frac{y \, z }{(z-4 y)  \cdot \,
 (y  \,+8 z y \, + 4 \cdot \, (4 y-1) \cdot \, z^2) }} }.
\end{eqnarray}
is {\em actually a solution of the three PDEs}, $ \, PDE_3^{(1)}$, $ \, PDE_3^{(2)}$ 
and $ \, PDE_2$. Unfortunaly, the other solutions are too complicated 
to be used to fabricate more general common solutions of the three PDEs.

However, the bivariate series $ \, T_2(z, y)$ can be written as a combination
of the solutions of $ \, L_5^{(z)}$.
Let us call $ \, S_1^{(z)}$, $ \, S_2^{(z)}$ and $ \, S_3^{(z)}$ the formal solutions
analytic at $z= \, 0$ of (respectively) the operators $\,L_1^{(2)} \cdot L_1^{(1)}$,
$\,L_1^{(3)} \cdot L_1^{(1)}$ and $\,L_2 \cdot L_1^{(1)}$.
The first terms of these solutions are given in \ref{combODEzT2}.

The bivariate series $ \, T_2(z, y)$ reads
\begin{eqnarray}
\hspace{-0.75in}&& \, \quad  \quad \, \, 
T_2(z, y) \, \,=\,\, \,\,
 C_1^{(z)}(y) \cdot S_1^{(z)} \,\, + C_2^{(z)}(y) \cdot S_2^{(z)}\,\, 
 + C_3^{(z)}(y) \cdot S_3^{(z)}, 
\end{eqnarray}
where the combination coefficients $ \, C_j^{(z)}(y)$ are given in  \ref{combODEzT2}.

\vskip 0.1cm
Similarly, one may consider the bivariate series $ \, T_2(z, y)$ as a combination
of the solutions of $N_5^{(y)}$.
With $ \, S_1^{(y)}$, $ \, S_2^{(y)}$ and $ \, S_3^{(y)}$ the formal solutions
analytic at $y=0$ of (respectively) the operators $\,N_1^{(2)} \cdot N_1^{(1)}$,
$\,N_1^{(3)} \cdot N_1^{(1)}$ and $\,N_2 \cdot N_1^{(1)}$ (see \ref{combODEyT2}),
the bivariate series $\, T_2(z, y)$ reads
\begin{eqnarray}
\hspace{-0.75in}&& \, \quad  \quad \, \, 
T_2(z, y) \,\,=\,\,\,\, C_1^{(y)}(z) \cdot S_1^{(y)} \,\,+ C_2^{(y)}(z) \cdot S_2^{(y)} 
\,\,+ C_3^{(y)}(z) \cdot S_3^{(y)}, 
\end{eqnarray}
where the combination coefficient $ \, C_j^{(y)}(z)$ are given in  \ref{combODEyT2}.

\vskip 0.1cm

Note that we have used for the solutions $S_j^{(z)}$ (resp. $S_j^{(y)}$) the
formal solutions of the corresponding operators since this is easier.
Otherwise a full closed expression for $T_2(z, y)$ is given in \ref{V2},
which is obtained by integration of the double integral.
One should note that the expression is a ``partition'' that does not reflect the
factorization of (e.g.) $L_5^{(z)}$.

\section{Partial differential equations for $ \, T_3(z,y)$}
\label{T3}

Similar calculations can be performed for the bivariate series $ \, T_3(z,y)$ 
corresponding to the expansion around $ \, (0, 0)$ of the integral 
(\ref{integralTdzy}) with $ \, d= \, 3$. In this instance, however, the 
creative telescoping method turned out to be too costly, and hence, 
all the PDEs presented below have been obtained by the guessing method.

We find that, for $Q= \, 3$ and $D= \, 3$, there is only one PDE (denoted $ \, PDE_3$) 
and for $Q = \, 4$, $D = \, 2$ there are two PDEs (called $PDE_4^{(1)}$, $ \, PDE_4^{(2)}$).
Here also, and similarly to $ \, T_2(z, y)$, 
both $ \, PDE_4^{(1)}$, $ \, PDE_4^{(2)}$ 
acting on the generic bivariate series (\ref{bivariateform}) generate 
the unique $ \, T_3(z, y)$, while $ \, PDE_3$ is sufficient to generate 
a unique solution that identifies with $ \, T_3(z, y)$.

As for the logarithmic solutions, there is no solution of the form (\ref{logForm})
for the system ($PDE_4^{(1)}$, $ \, PDE_4^{(2)}$).
However, and similarly to what happened for $ \, T_2(z,y)$, the number of 
logarithmic solutions for $ \, PDE_3$ depends on the value of $ \, \mu$ 
in the combination (\ref{logFormmu}).
For $\mu= \, 1$ and $\mu= \, 1/2$, there is non finite number of such solutions
(we reached $n= \, 17$ in our calculations).

For generic values of $\mu  \,\ne \, 1, \, 1/2$, one obtains three solutions, 
the bivariate series $ \,T_3(z, y)$ and the logarithmic solutions
\begin{eqnarray}
\hspace{-0.75in}&& \, \quad  \quad 
T_3(z,y) \cdot \, (\ln(z) \, - \mu \cdot \, \ln(y))^2 \, \, \, \,
+ T_3^{(1)} \cdot  \, ( \ln(z) - \mu \cdot \, \ln(y)) 
 \,\, \, \, +  \,  \,T_3^{(0)},
 \nonumber \\
\hspace{-0.75in}&& \, \quad  \quad  
T_3(z,y) \cdot \, (\ln(z)  \,- \mu \cdot \, \ln(y)) 
 \,\, \, + {1 \over 2} \, T_3^{(1)},  
\end{eqnarray}
where:
\begin{eqnarray}
\hspace{-0.95in}&&  \quad  \quad 
T_3^{(1)} \,  \,=\, \, \, \,  \,
4 \cdot \, (1 \, -4\,\mu) \cdot  \, z \, \,\, \,
  -4 \cdot \, (7\,\mu-1) \cdot  \, y \,\,\,   \,
-2 \cdot  \, (24\,\mu-17) \cdot \, {z}^{2}
 \nonumber \\
\hspace{-0.95in}&& \,  \quad  \quad \quad \quad 
+4 \cdot \, (13-40\,\mu) \cdot  \, y z \,\,
 -6 \cdot \, (87\,\mu-7)\cdot  {y}^{2} \,\,  \,  \,  + \, \cdots
\end{eqnarray}
\begin{eqnarray}
\hspace{-0.95in}&&   \quad  \quad 
T_3^{(0)} \,  \,=\, \,\,   \,
8 \cdot \,\mu \cdot \, (1-5\,\mu) \cdot \, z \,\, \, 
 +2 \cdot \, (8\,\mu-11)\cdot \,  y \, \,  \,
\, -2 \cdot \, (3\,\mu-2+31\,{\mu}^{2})\cdot \,  {z}^{2}
  \nonumber \\
\hspace{-0.95in}&& \,   \quad  \quad \quad 
-{4 \over 3} \cdot \, (57\,{\mu}^{2}+41-9\,\mu) \cdot \,  y z \, \, \,  \, 
+{1 \over 2} \cdot \, (700\,\mu +392\,{\mu}^{2}-831) \cdot  \, {y}^{2} 
 \,  \, \,+  \,\cdots
\end{eqnarray}

\subsection{Decoupled linear differential equations for $\, T_3(z,y)$}
\label{decoupledT3}

For the PDE system $\bigl\{PDE_3,\,PDE_4^{(1)},\,PDE_4^{(2)}\bigr\}$
annihilating $ \, T_3(z,y)$, we used Buchberger's algorithm as implemented in the
HolonomicFunctions program~\cite{koutschan-package}
and obtained immediately a Gr\"obner basis (given in electronic form in~\cite{koutschan-webpage}).
It allows us to derive two (order-nine) ordinary differential 
equations\footnote[2]{Do not confuse the labels of some factors with those
occurring for $ \, T_2(z,y)$.}
for $ \, T_3(z, y)$, one involving only $ \, D_z$,  the other one only $ \, D_y$: 
\begin{eqnarray}
\label{defL9z}
\hspace{-0.95in}&& \quad \quad \,  \, \, \,
L_9^{(z)}\,\,   = \, \, \, \sum_{n=0}^{9} \, P_n(z, y) \cdot \, D_z^n,
 \qquad \, \, 
N_9^{(y)}\,\,   = \, \, \, \sum_{n=0}^{9} \, Q_n(z, y) \cdot \, D_y^n.
\end{eqnarray}
One factorization of $ \, L_9^{(z)}$ 
reads\footnote[1]{The full factorization of $ \, L_9^{(z)}$
as a {\em direct sum} is  $\, L_9^{(z)} =$
$ {\tilde L}_3 \cdot L_2 \oplus L_1^{(1)} \cdot L_2 
\oplus {\tilde L}_1^{(2)} \cdot L_2 \oplus 
{\tilde L}_1^{(3)} \cdot L_2 \oplus {\tilde L}_1^{(4)} \cdot L_2$.} 
(the indices denote orders):
\begin{eqnarray}
\label{L9z}
\hspace{-0.95in}&& \quad \qquad   \qquad
L_9^{(z)} \, \,  =\, \,  \, 
L_3 \cdot L_1^{(4)} \cdot L_1^{(3)} \cdot L_1^{(2)} \cdot L_1^{(1)} \cdot L_2.
\end{eqnarray}
The similar factorization of $ \, N_9^{(y)}$ reads:
\begin{eqnarray}
\label{N9y}
\hspace{-0.95in}&& \quad \qquad   \qquad
N_9^{(y)} \,  \, =\,\,  \, 
N_3 \cdot N_1^{(4)} \cdot N_1^{(3)} \cdot N_1^{(2)} \cdot N_1^{(1)} \cdot N_2. 
\end{eqnarray}
The two order-two operators $\, L_2$ and $\, N_2$  are 
given in \ref{rightmost}.
They are self-adjoint, up to a conjugation by their Wronskians 
$\, W(L_2)$ and $\, W(N_2)$ (see \ref{rightmost}):
\begin{eqnarray}
\label{upto}
\hspace{-0.95in}&& 
L_2 \cdot \, W(L_2)  \,   =\,\, W(L_2)  \cdot \, adjoint(L_2), \quad 
N_2 \cdot \, W(N_2)  \,   =\,\, W(N_2)  \cdot \, adjoint(N_2).
\end{eqnarray}

\vskip 0.1cm

We have been able to find one solution for $\, L_9^{(z)}$ (and $ \, N_9^{(y)}$).  
Defining the hypergeometric function
\begin{eqnarray}
\label{Szy}
\hspace{-0.95in}&& \quad \qquad   
S_{zy}\, \,=\,\,\,
 {\frac{\sqrt{y\, z}}{\sqrt{P_{zy}} }} \cdot \, 
_{2}F_1\left( [{{1} \over {4}}, \,{{3} \over {4}}],[1], \, 
{\frac{64 \cdot \,   y \,z  \cdot \, 
(z-3 y)^3 \cdot \,(1+4 z)^2}{P_{zy}^2}} \right), 
\end{eqnarray}
where
\begin{eqnarray}
\label{Pzy}
\hspace{-0.95in}&&  
P_{zy} \,=\,
1728\,{y}^{2}{z}^{3}-432\,y{z}^{3}+16\,{z}^{3}+864\,{z}^{2}{y}^{2}
-72\,{z}^{2}y+108\,z{y}^{2}+zy-4\,{y}^{2}, 
\end{eqnarray}
one checks that $ \,  \, S_{zy}$ 
is solution of the {\em two
most right order-two operators} $\,L_2$ and $\,N_2$
\begin{eqnarray}
\label{L2N2}
\hspace{-0.95in}&& \quad \quad \qquad   \qquad
L_2 (S_{zy}) \,=\,\,0,
 \qquad \, \, 
N_2 (S_{zy}) \,=\,\,0.
\end{eqnarray}
As was seen for the operators $\,L_5^{(z)}$ and $\,N_5^{(y)}$ 
corresponding to $\,T_2(z,y)$,
with the solution (\ref{solPDET2}), 
one can check that the solution (\ref{Szy}) of 
$\, L_9^{(z)}$ (and $ \, N_9^{(y)}$), is one solution to the whole PDE system:
\begin{eqnarray}
\label{PDE344}
\hspace{-0.95in}&& \quad \quad
PDE_3 (S_{zy}) \,=\,\,0, \quad \quad
PDE_4^{(1)} (S_{zy}) \,=\,\,0, \quad \quad
PDE_4^{(2)} (S_{zy}) \,=\,\,0.
\end{eqnarray}

\vskip 0.2cm

This solution (\ref{Szy}) of the whole PDE system is, in fact, 
quite remarkable. It corresponds to a 
{\em modular form}~\cite{Maier,CalabiYauIsing,IsingModularForms}. In
order to see this modular form structure, let us recall various 
(non trivial) identities on hypergeometric functions. 

\vskip 0.2cm

The use of the identity 
\begin{eqnarray}
\label{identi3over4}
\hspace{-0.95in}&& \, \,  
_{2}F_1\left( [{{1} \over {4}}, \, {{3} \over {4}}],[1], \, X  \right)
 \,= \,\,\,  
{\frac{1}{(1+3 X)^{1/4}}} \cdot \, 
_{2}F_1\left( [{{1} \over {12}}, \, {{5} \over {12}}],[1],\, 
{\frac{27 X \cdot \, (1-X)^2}{(1+3X)^3}}   \right), 
\end{eqnarray}
together with the identity 
\begin{eqnarray}
\label{identi3over4a}
\hspace{-0.95in}&&  \, \,  
_{2}F_1\left( [{{1} \over {4}}, \, {{3} \over {4}}],[1], \, X  \right) \,= \,\,\,  
\Bigl({\frac{4}{4-3 X}}\Bigr)^{1/4} \cdot \, 
_{2}F_1\left( [{{1} \over {12}}, \, {{5} \over {12}}],[1], \, 
{\frac{27 X^2 \cdot \, (X-1)}{(3X -4)^3}}   \right), 
\end{eqnarray}
implies the following identity on {\em the same}
 hypergeometric function 
\begin{eqnarray}
\label{identi1over12}
\hspace{-0.95in}&& \quad 
_{2}F_1\left( [{{1} \over {12}}, \, {{5} \over {12}}],[1],\,A_1   \right) 
\,= \,\,
\sqrt{2} \cdot \, \left( {\frac{1+3 X}{4-3 X}} \right)^{1/4} \cdot \, 
_{2}F_1\left( [{{1} \over {12}}, \, {{5} \over {12}}],[1],\,A_2   \right), 
\end{eqnarray}
with the 
{\em two different arguments}\footnote[1]{Such non-trivial identity 
(\ref{identi1over12}) is characteristic of 
{\em modular forms}~\cite{Maier,CalabiYauIsing,IsingModularForms}.}
\begin{eqnarray}
\label{A1A2}
\hspace{-0.95in}&&  
A_1(X) \,= \, \,  {\frac{27 \cdot \, X \cdot \, (1-X)^2}{(1 \, +3X)^3}}, \, \,  \quad 
A_2(X)\,= \,  \, {\frac{27 \cdot \, X^2 \cdot \, (X-1)}{(3X \, -4)^3}}\,= \,  \, 
A_1(1\, -X). 
\end{eqnarray}
This enables to rewrite the solution $ \, S_{zy}$ of (\ref{PDE344}),
where  $\, S_{zy}$ is given in (\ref{Szy}), 
as a $\,  _{2}F_1$ hypergeometric function with 
{\em two different pullbacks}, namely 
$\, A_1$ and $ \, A_2$ given by (\ref{A1A2}) where $\, X$ is given by 
(with $\, P_{zy}$ given by (\ref{Pzy})): 
\begin{eqnarray}
\label{X}
\hspace{-0.95in}&& \quad \qquad   \qquad
X \, \, = \, \, \, 
{\frac{64 \cdot \,  y \, z \cdot \, (z-3 y)^3 \cdot \, (1+4 z)^2}{P_{zy}^2}}, 
\end{eqnarray}
and where $\, 1\, -X \, $ reads:
\begin{eqnarray}
\label{1moinsX}
\hspace{-0.95in}&& 
 {\frac{ (144\,{y}^{2}{z}^{2}+96\,{y}^{2}z
-40\,y{z}^{2}+16\,{y}^{2}-8\,yz+{z}^{2})
\,   (144\,y{z}^{2}+24\,yz-16\,{z}^{2}+y)^{2}
}{P_{zy}^2}}. 
\end{eqnarray}

This shows that the solution $\,  S_{zy}$, given in (\ref{Szy}), 
corresponds to a {\em modular form}, seen as 
a function of  $ \, z$, or seen as a function of $\, y$. 

\vskip .1cm 

The two pullbacks $ \, A_1$ and $ \, A_2$ are lying on the 
 algebraic genus zero {\em modular curve}
\begin{eqnarray}
\label{modularA1A2}
\hspace{-0.95in}&& \quad \quad  
 1953125 \cdot\, A_1^3\, A_2^3
\, \,\,  -187500 \,\cdot \, A_1^2\, A_2^2 \cdot \, (A_1+A_2) \, 
\nonumber \\
\hspace{-0.95in}&& \quad \quad  \quad \quad  
\, +375 \cdot \, A_1\, A_2 \cdot \,
 (16\, A_1^2 -4027\, A_1\, A_2 +16\, A_2^2)
 \\
\hspace{-0.95in}&& \quad \quad  \quad \quad  
\, -64 \cdot \, (A_1+A_2) \cdot \,
 (A_1^2+1487\, A_1\, A_2+A_2^2) \,\,\,
+110592\, A_1\, A_2
\,\, = \,\,\,\,0. \nonumber
\end{eqnarray}
If one introduces $\, Z$ such that $\, X \, = \, \, Z/(Z+64)$, one 
can see clearly that this {\em modular equation}\footnote[2]{G.S. Joyce 
already noticed the emergence of {\em modular equations} on lattice Green 
functions~\cite{Joyce}.
} (\ref{modularA1A2}) 
is the same as the one corresponding to the {\em fundamental
modular curve} $\, X_0$, associated with the Landen transformation,
and its well-known Hauptmodul rational 
parametrization~\cite{Maier,CalabiYauIsing,IsingModularForms}:
\begin{eqnarray}
\label{modularA1A2bis}
 A_1 \,\, = \, \, \, {{1728 \cdot \, Z } \over {(Z+16)^3 }}, \qquad  \, \, 
A_2 \,\, = \, \, \, {{1728 \cdot \, Z^2} \over {(Z+256)^3 }}.
\end{eqnarray}

\vskip .2cm 

{\bf Remark:} We have a quite remarkable result for the 
 $\, X \, = \, const. \, \, $ foliation. 
When $\, X$ is a constant, one finds that the curves 
$\, X \, = \, const.$,  {\em are genus zero curves}. 

\vskip .1cm

\section{Remarks and comments}
\label{remcom}

We give here some miscellaneous remarks on the calculations presented
in the previous sections.

\vskip 0.1cm

\noindent
{\bf Remark 1:}
The system of recurrence equations for $\, t_2(n,j)$ can be obtained 
by the creative telescoping method~\cite{zeilberger-ct,koutschan-survey}. 
For higher $d$, the computations get too heavy.
But with the guessing method, the recurrences for $d=5$ can be 
reached and this may yield an efficient implementation, since 
one could take $\, t _5(n, j)$ as initial values instead of 
$\, t_2(n, j)$ in~(\ref{Tdnj}).

\vskip 0.1cm

\noindent
{\bf Remark 2:} 
In our calculations, (Sections~\ref{sec:odes} and~\ref{computdetails}), 
we have experienced that we do not gain anything by doing the computation 
modulo primes, and then using chinese remaindering to construct 
the true result in~$\mathbb{Z}$. The reason is that here we do not 
encounter an intermediate expression swell, but 
rather the fact that the largest integers that occur during the computation are 
basically those that are given as the final result (when this is considered to 
be the list of Taylor coefficients).  Of course, we are mostly interested in the 
linear differential operator, which, itself, has much smaller 
integer coefficients. Therefore the natural strategy would be to compute the 
Taylor coefficients modulo prime, and guess the operator modulo prime, and, 
only after this is done for sufficiently many primes, use chinese remaindering 
and rational reconstruction to get the true operator. Unfortunately, the operator 
also has quite large integers in its coefficients so that this strategy is 
unfavorable. However, we can still use homomorphic images for the purpose
of prediction, e.g. how many terms are required for guessing the linear
differential operator.

\vskip 0.1cm

As an example, in the case ($d=\, 11$, $20$ processes, $2464$ terms), the timing
is $18$ days in exact arithmetic calculations. When we compute modulo the 
prime $2^{31}-1$, our implementation needs $58$ hours, but the rational reconstruction 
of the linear differential operator requires $185$ primes of this size.

\vskip 0.1cm

\noindent
{\bf Remark 3:}
With the  emergence of the algebraic solution (\ref{solPDET2})
for the system of $\, d= \, 2$ PDEs, 
 and the  emergence of the modular form solution (\ref{Szy})
 for the system of $\, d= \, 3$ PDEs, it is tempting to conjecture that
similar solutions of two variables exist for all the system of PDEs
for arbitrary value of $\, d$.

\vskip 0.1cm

\noindent
{\bf Remark 4:}
For one complex variable, the  holonomic (or D-finite~\cite{Lipshitz}) functions 
are solutions of linear ODEs with polynomial coefficients in the complex variable. 
The singularities (and apparent singularities) can be seen immediately as 
solutions of the head polynomial coefficient of the linear ODE.
For partial differential equations system annihilating holonomic functions of 
several complex variables, the singular manifolds would be too complex or simply
could not be well defined. 
By considering several Picard-Fuchs systems of two-variables ``associated'' to 
Calabi-Yau ODEs\footnote[1]{Along the line of the relation between lattice 
Green functions and Calabi-Yau ODEs see for instance~\cite{guttmann-2009}.}, we 
showed~\cite{2012-pde-to-ode-anisIsing} that D-finite (holonomic) functions 
are actually a good framework for actually finding properly the singular 
manifolds. The singular algebraic varieties for some $ \, T_d (z,y)$ 
are given in \ref{singTdzy}.

\vskip 0.1cm

\section{Conclusion}
\label{Conclusion}
A {\em recursive method} has been introduced in~\cite{2015-LGF-fcc7}
 to generate the expansion of the lattice Green function of the 
$d$-dimensional face-centred cubic lattice. 
The method has been used to generate many coefficients for $\,d=\,7$
and the corresponding linear differential equation has been 
obtained~\cite{2015-LGF-fcc7}.

We have shown, here, the strength and the limit of this  recursive method.
Some observations on the recursive method allow us to improve the 
computations and produce the series up to $d=12$. The corresponding 
linear differential equations have been obtained (available online~\cite{koutschan-webpage}) and show that the
pattern (order, singularities, differential Galois group) seen for the
lower $d$'s continues, as discussed in Sections~\ref{sec:odes} and~\ref{sec:galois}.

In the recursive method, a two-dimensional array $ \, t_d(n, j)$, defined in (\ref{Tdnj})--(\ref{T2nj2}), is computed 
where only the coefficients $ \, t_d(n, 0)$ correspond to the expansion
of the lattice Green function. The two-dimensional array $ \, t_d(n, j)$  
gives the expansion of a ``lattice'' Green function $\,T_d(z,y)$ that depends on two
variables. These $D$-finite bivariate series are studied,
in Sections~\ref{partialdiff} and~\ref{ODET2} for $d=\,2$ and in Section~\ref{T3} for $d=\,3$,
and the differential equations they are solution of, are addressed.

We have been able to produce some solutions of the partial differential 
equations annihilating the bivariate series $ \, T_d(z, y)$.
In Section~\ref{T3}
 a remarkable {\em modular form} solution emerged for $d= \, 3$.
The corresponding Hauptmodul pullback is a simple rational function
of $\, y$ and $\, z$.
In terms of this Hauptmodul, the $(y, \, z)$-plane is a 
foliation of rational curves. 
Such kind of results are clearly a strong incentive to
generalize the search of solutions of the $\, D$-finite systems 
corresponding to higher dimensions~$\, d$.

\vskip .5cm

{\bf Acknowledgments:} One of us (JMM) would like to thank
A.J. Guttmann for so many friendly and fruitful holonomic, 
lattice Green, enumerative combinatorics discussions
during the last two decades.
One of us (JMM) would like to thank
 G. Christol and J-A. Weil for fruitful discussions 
respectively on diagonals of rational functions,
and $\, D$-finite systems of PDEs.
This work has been performed without
 any support of the ANR, the ERC, the MAE or any PES of the CNRS. 
But one of us (CK) was supported by the Austrian Science Fund (FWF): W1214.

\appendix

\section{Singularities of the ODEs for $d= \, 8, \, 9, \, 10,\,  11,\,  12$}
\label{sing891011}

The singularities, occurring at the head derivative of $\, G_{14}^{8Dfcc}$, are
$\, {x}^{11} S_8(x) \cdot \, P_{14}(x) $.
The roots of the degree-95 polynomial $ \, P_{14}(x)$ are {\em apparent singularities}.
The polynomial $ \, S_8(x)$ corresponding to the finite singularities reads:
\begin{eqnarray}
\hspace{-0.95in}&&  \quad     
 S_8(x)\,  \,=\, \,\, 
 (14+x)^2 \, (x-1)  \, (x-7)  \, (x-2)  
 \, (6+x)  \, (7+x)   \, (20+x) 
 \left( 28+x \right)
 \nonumber \\
\hspace{-0.95in}&&  \quad   \quad  \quad             
 (48+x)  \, (21+2\,x) 
 \, (4+3\,x)  \, (28+3\,x)  \, (32+3\, x) 
 \, (16+5\,x)  \, (28+5\,x)
  \nonumber \\
\hspace{-0.95in}&&  \quad  \quad  \quad       
  (28+11\,x)  \, (112+11\,x)  \, (224+13\,x) 
 \, (112+19\,x).  
\end{eqnarray}

\vskip 0.1cm

The singularities of $ \, G_{18}^{9Dfcc}$ are $\, {x}^{14} \cdot \, S_9(x) \, P_{18}(x) $.
The roots of the degree-133 polynomial $\, P_{18}(x)$ are {\em apparent singularities}.
The polynomial $ \, S_9(x)$ corresponding to the finite singularities reads:
\begin{eqnarray}
\hspace{-0.95in}&&  \quad  
     S_9(x)\,  \,=\,\, \, 
 (x+9)^4
 \, (7\,x+9)  \, (4\,x+9)  \, (x+3) 
 \, (2\,x+9)  \, (7\,x+36)  \, (5\,x+27)    \, (x+12)
 \nonumber \\
 \hspace{-0.95in}&&  \quad   \quad \quad      
 (2\,x+27)  \, (5\,x+72)  \, (x+15)  \, (x+18) 
 \, (2\,x+45)  \, (x+27)  \, (x+36)  \, (x+63) 
\nonumber \\
 \hspace{-0.95in}&&  \quad  \quad   \quad     
 (2\,x-9)  \, (5\,x-9)  \, (x-1).  
\end{eqnarray}

\vskip 0.1cm

The singularities of $ \, G_{22}^{10Dfcc}$ are 
$\, {x}^{18} \cdot \,  S_{10}(x) \cdot  \, P_{22}(x) $.
The roots of the degree-252 polynomial $ \, P_{22}(x)$ are {\em apparent singularities}.
The polynomial $\, S_{10}(x)$, corresponding to the finite singularities, reads: 
\begin{eqnarray}
\hspace{-0.95in}&&   
S_{10}(x) \,=\, \,  (15+x)^{2} \, (x-1)  \, (4\,x+75)  
\, (19\,x+540)  \, (x-15)  \, (4\,x+15)  \, (x+80)
  \nonumber \\
\hspace{-0.95in}&&  \quad \, \, \, \,   \,  
(2\,x+45)  \, (22\,x+45)  \, (2\,x+25) 
 \, (7\,x+20)  \, (7\,x+120) 
 \, (2\,x+15)  \, (23\,x+180)   
 \nonumber \\
\hspace{-0.95in}&&  \quad \, \,  \, \,  \,  
(13\,x+60) 
 \, (29\,x+360)  \, (17\,x+135)  \, (x+45) 
 \, (4\,x+45)  \, (x+35)  \, (3\,x-5)
  \\
\hspace{-0.95in}&&  \quad  \, \, \, \,  \,  
(4\,x+5)  \, (11\,x+135)  
\, (x+20)  \, (13\,x-45)  \, (x+12) 
 \, (x+9)  \, (x+8)  \, (x+5). 
 \nonumber
\end{eqnarray}

\vskip 0.1cm

The singularities of $ \, G_{27}^{11Dfcc}$ are $\, {x}^{22} S_{11}(x) \, P_{27}(x) $.
The roots of the degree-352 polynomial $\, P_{27}(x)$ are {\em apparent singularities}.
The polynomial $\, S_{11}(x)$, corresponding to the finite singularities, reads: 
\begin{eqnarray}
\hspace{-0.95in}&&  \quad 
S_{11}(x) \, \,=\, \, \, 
 (x+11)^{6} \, (55+x)^{2}
\, (x-1)  \, (8\,x+55)  \, (29\,x+55)  
\, (4\,x+55)  \, (2\,x+55)
  \nonumber \\
\hspace{-0.95in}&&  \quad  \quad \, \,  \, \,
(4\,x+11)  \, (7\,x+165)  \, (7\,x-55)
  \, (2\,x+33)  \, (17\,x+55) 
 \, (x+44)  \, (13\,x+275) 
   \nonumber \\
\hspace{-0.95in}&&  \quad  \quad \, \,  \, \,
(3\,x+55)  \, (7\,x-11) 
 \, (13\,x+55)  \, (7\,x+110)  \, (x+35)  
\, (3\,x+22)  \, (x+99)   
  \nonumber \\
\hspace{-0.95in}&&  \quad  \quad \, \,  \, \,
(19\,x-55)  \, (7\,x+33)  \, (9\,x+11)
  \, (x+15)  \, (9\,x+55)  \, (17\,x+275)  \, (3\,x+77) 
   \nonumber \\
\hspace{-0.95in}&&  \quad  \quad  \, \,  \, \,
(23\,x+165).
\end{eqnarray}

\vskip 0.1cm

The singularities of $ \, G_{32}^{12Dfcc}$ are 
$\, {x}^{27} S_{12}(x) \, P_{32}(x) $.
The roots of the degree-580 polynomial $\, P_{32}(x)$ 
are {\em apparent singularities}.
The polynomial $\, S_{12}(x)$, corresponding to the finite 
singularities, reads: 
\begin{eqnarray}
\hspace{-0.95in}&&  \quad 
S_{12}(x) \, \,=\, \, \, 
 \, (2\,x+33)^{2} \, (43\,x+264)  \, (5\,x+6) 
 \, (37\,x+66)  \, (3\,x+8)  \, (23\,x+66)
 \nonumber \\
\hspace{-0.95in}&&  \quad  \quad \, \, 
 \, (67\,x+264)  \, (2\,x+9) 
 \, (13\,x+66)  \, (5\,x+33)  \, (7\,x+48) 
 \, (7\,x+66)  \, (9\,x+88) 
\nonumber \\
\hspace{-0.95in}&&  \quad  \quad \, \, 
 \, (10\,x+99)  \, (53\,x+528)  \, (x+10) 
 \, (x+11)  \, (5\,x+66)  \, (19\,x+264) 
 \, (7\,x+99)
  \nonumber \\
\hspace{-0.95in}&&  \quad  \quad \, \, 
 \, (37\,x+528)  \, (23\,x+330) 
 \, (5\,x+72)  \, (29\,x+528)  \, (17\,x+330)  \, (13\,x+264) 
 \nonumber \\
\hspace{-0.95in}&&  \quad  \quad \, \, 
  \, (x+21)  \, (x+22)  \, (31\,x+792)  \, (8\,x+231)  
\, (x+32)  \, (x+33)  \, (25\,x+1056) 
 \nonumber \\
\hspace{-0.95in}&&  \quad  \quad \, \, 
 \, (x+54)  \, (x+66)  \, (x+120) 
 \, (x-33)  \, (2\,x-11)  \nonumber \\
\hspace{-0.95in}&&  \quad  \quad \, (13\,x-33) 
 \, (2\,x-3)  \, (x-1).
\end{eqnarray}

\section{Gr\"obner basis basics}
\label{groebner}

The theory of Gr\"obner bases has been initiated by Bruno Buchberger in his
Ph.D. thesis~\cite{buchberger-1965} in 1965. While originally it was
formulated for commutative multivariate polynomial rings, we are interested in
its generalization to noncommutative rings. Here we can only mention a few key
facts that are important for the kind of applications that we have in mind. A 
very instructive introduction to Gr\"obner bases is given 
in~\cite{CoxLittleOShea92}.

Let $ \, D_z$ denote the operator $ \, \partial/\partial z$. The motivation 
for using operator notation is that it turns ODEs and PDEs into (univariate
resp. multivariate) polynomials. For example the PDEs appearing in Sections
\ref{partialdiff}--\ref{T3} can be represented by polynomials in the ring
$\mathbb{O}=\mathbb{C}(z,y)[D_z,D_y]$, i.e., the ring of partial differential
operators in~$z$ and~$y$ (it is an instance of an {\em Ore algebra}). Note
that this ring is not commutative, because of the Leibniz rule
$D_z\,a=a\,D_z+\partial a/\partial z$ for all $a\in\mathbb{C}(z,y)$.

Let $f$ be a power series (or some other kind of ``function''); we define
\[
  \ann_{\mathbb{O}}(f) = \bigl\{ P\in\mathbb{O} \mathrel{\big|} P(f) = 0 \bigr\},
\]
called the {\em annihilating ideal} of~$f$. It can be easily seen that this
set is indeed a left ideal in~$\mathbb{O}$, as for example, the
left-multiplication of $P\in\mathbb{O}$ by~$D_z$ corresponds to
differentiating the differential equation represented by~$P$ with respect
to~$z$. Since univariate polynomial rings are principal ideal domains,
$\ann(f)$ is generated by a single element if we consider only one derivation;
this unique generator corresponds to the minimal-order ODE.
In fact the Gr\"obner basis computation specializes to the
\emph{greatest common right divisor} in this setting. Of course, in the
case of PDEs, an annihilating ideal in general is generated by several
operators.

We need some notion of {\em leading term} for PDEs (in the case of ODEs
it is clear). For this purpose one imposes a total order on the monomials in the
ring under consideration, that is compatible with multiplication and that has
$1$ as the smallest monomial; such an ordering is called a {\em monomial order}.
For example, the degree-lexicographic order on the ring 
$\,\mathbb{C}(z,y)[D_z,D_y]\,$
with $\, D_y \prec D_z \, $ is defined by
\begin{eqnarray}
\hspace{-0.95in}&&  \quad  \quad 
  D_z^i \, D_y^j\, \prec \,D_z^k \, D_y^\ell \, \,  
  \, \>\iff\>\, \, \,  
 i \,+j \,<\, k \,+\ell \, \lor\, \bigl(i \,+j=\,k \,+\ell \,  \land \, i \,<\,k \, \bigr).
\end{eqnarray}

Using this notion of leading term, it is straightforward to define a
multivariate polynomial division (called {\em reduction}) of $P\in\mathbb{O}$
by some $Q_1,\ldots,Q_r\in\mathbb{O}$. It works by subtracting, in each step,
a suitable multiple of some~$Q_i$ such that the leading term of the dividend
vanishes. In some steps of this process one may have the choice between
several of the $Q_i$, and this has the consequence that the remainder of the
multivariate polynomial division is not unique in general.  Now, if $I$ is an
ideal, then a set~$G$ of generators of~$I$ is called a {\em Gr\"obner basis}
if for each polynomial the remainder of the division by~$G$ is unique. In
particular, we have that the division of $P$ by $G$ has remainder~$0$ if and
only if $P\in I$; this property allows to decide the {\em ideal membership
  problem}. Gr\"obner bases are also a powerful tool for elimination purposes,
i.e., for finding elements in an ideal that do not depend on some of the
variables of the polynomial ring. There are several algorithms to compute,
from an arbitrary set of generators of an ideal, a Gr\"obner basis, the most
classic one being Buchberger's algorithm~\cite{buchberger-1965}.

\section{The factorization of $\, L_5^{(z)}$ and $\, L_5^{(y)}$ for $\, T_2(z, y)$}
\label{factorizL5zN5y}

The factors in the decomposition (\ref{directsumdecompL}) of $\, L_5^{(z)}$ read  
\begin{eqnarray}
\hspace{-0.95in}&&  \quad   
L_1^{(1)} \, \,=\,\, \, \,
D_z\,\,\, +2 \cdot \,{\frac {8\,{z}^{3}y -16\,{z}^{2}{y}^{2} +6\,{z}^{2}y
-2\,{z}^{3}+{y}^{2}}{ (16\,{z}^{2}y +y+8\,zy-4\,{z}^{2}) 
 \, (z-4\,y) \cdot \,  z}}, 
\end{eqnarray}
\begin{eqnarray}
\hspace{-0.95in}&&  \quad  
L_1^{(2)} \,\,=\,\,\,\,
D_z\,\,\, +2 \cdot \,{\frac {32\,{z}^{3}y -8\,{z}^{3} -96\,{z}^{2}{y}^{2} +36\,{z}^{2}y
+zy-32\,{y}^{2}z -2\,{y}^{2}}{ (16\,{z}^{2}y +y +8\,zy -4\,{z}^{2})
 \, (z-4\,y) \cdot \,z}}, 
\end{eqnarray}
\begin{eqnarray}
\hspace{-0.95in}&&  \quad  
L_1^{(3)}\, \,=\,\,\, \,   L_1^{(2)} \,\,\, - {1 \over z}, 
\end{eqnarray}
and
\begin{eqnarray}
\hspace{-0.95in}&&  \quad  \quad \quad \quad \quad \quad 
L_2 \,\,=\,\,\,\, D_z^2 \,\,\,\,
+ {\frac{p_1(z,y)}{p_2(z,y)}} \cdot \, D_z \,\,\, \,+ {\frac{p_0(z,y)}{p_2(z,y)}}, 
\end{eqnarray}
where:
\begin{eqnarray}
\hspace{-0.95in}&&   
p_2(z,y) \,\,=\,\,\,\,
z \cdot \, (4\,z-1) \cdot  \, (4\,z+1)  
\cdot \, (y+8\,yz+16\,y{z}^{2}-4\,{z}^{2})  \cdot \, (z-4\,y) 
 \\
\hspace{-0.95in}&&  \quad   \quad \, \,\,
  (256\,{z}^{4}{y}^{2}-64\,{z}^{4}y-128\,{z}^{3}{y}^{2}
+4\,{z}^{3}+64\,{z}^{2}{y}^{2}-
20\,y{z}^{2}+40\,{y}^{2}z -2\,yz+3\,{y}^{2}), 
\nonumber 
\end{eqnarray}
\begin{eqnarray}
\hspace{-0.95in}&&  
p_1(z,y) \,\,=\,\,\,\,
2\,z \cdot \, \Bigl( {y}^{3}\,+16\,{y}^{4}\,-3200\,{z}^{4}{y}^{2}\,
-208\,{z}^{4}y\,+816\,{z}^{3}{y}^{2}+84\,{z}^{2}{y}^{2}+{y}^{2}z
 \nonumber \\
\hspace{-0.95in}&&  \quad    \quad \,
-18\,{z}^{3}y\,+576\,y{z}^{5}\,-40704\,{z}^{5}{y}^{2}
-46080\,{z}^{6}{y}^{2}\,+8704\,y{z}^{6}\, +3584\,{z}^{7}y 
\nonumber \\
\hspace{-0.95in}&&  \quad   \quad \,
-98304\,{z}^{8}{y}^{2}\, +12288\,{z}^{8}y\,+196608\,{z}^{8}{y}^{3}\,
-217088\,{z}^{4}{y}^{4}\,+181248\,{z}^{5}{y}^{3}
 \nonumber \\
\hspace{-0.95in}&&  \quad   \quad \,
+212992\,{z}^{7}{y}^{3}\,-24576\,{z}^{6}{y}^{3}\,+196608\,{z}^{6}{y}^{4}
-524288\,{z}^{7}{y}^{4}\,-98304\,{z}^{5}{y}^{4}
 \nonumber \\
\hspace{-0.95in}&&  \quad  \quad \,
-63488\,{z}^{3}{y}^{4}\,+89088\,{z}^{4}{y}^{3}-
20480\,{z}^{7}{y}^{2}\,-1440\,{z}^{2}{y}^{3} \,+6080\,{z}^{3}{y}^{3}
 \nonumber \\
\hspace{-0.95in}&&  \quad   \quad \,
-116\,z{y}^{3}\,-3840\,{z}^{2}{y}^{4}\,+384\,z{y}^{4} 
+24\,{z}^{5}\,-896\,{z}^{7} \Bigr), 
\end{eqnarray}
\begin{eqnarray}
\hspace{-0.95in}&&  \quad 
p_0(z,y) \,\,=\,\,\,
144\,{y}^{4}\,-7296\,{z}^{4}{y}^{2}\,-112\,{z}^{4}y\, +528\,{z}^{3}{y}^{2}\,
+124\,{z}^{2}{y}^{2}\,-2\,{y}^{2}z\,
 \nonumber \\
\hspace{-0.95in}&&  \qquad \quad \,
-28\,{z}^{3}y\, +1600\,y{z}^{5}\,-80896\,{z}^{5}{y}^{2}\, 
-107520\,{z}^{6}{y}^{2}\,+18176\,y{z}^{6}\,+8192\,{z}^{7}y 
\nonumber \\
\hspace{-0.95in}&& \qquad \quad \,
-196608\,{z}^{8}{y}^{2}\, +24576\,{z}^{8}y\,+393216\,{z}^{8}{y}^{3}
-315392\,{z}^{4}{y}^{4}\, +347136\,{z}^{5}{y}^{3}
 \nonumber \\
\hspace{-0.95in}&& \qquad \quad \,
+49152\,{z}^{7}{y}^{3}\,-8192\,{z}^{6}{y}^{3}\,+458752\,{z}^{6}{y}^{4}
-524288\,{z}^{7}{y}^{4}\,-229376\,{z}^{5}{y}^{4} 
\nonumber \\
\hspace{-0.95in}&&  \qquad \quad \,
-88064\,{z}^{3}{y}^{4}\,+150016\,{z}^{4}{y}^{3}\,+20480\,{z}^{7}{y}^{2}
-736\,{z}^{2}{y}^{3}\,+11840\,{z}^{3}{y}^{3}
 \nonumber \\
\hspace{-0.95in}&& \qquad \quad \,
-188\,z{y}^{3}-8960\,{z}^{2}{y}^{4}+384\,z{y}^{4}
+16\,{z}^{5}-2048\,{z}^{7}. 
\end{eqnarray}

\vskip 0.1cm

The factors in the decomposition  (\ref{N5blabla}) of $ \, N_5^{(y)}$ read   
\begin{eqnarray}
\hspace{-0.95in}&& \qquad
N_1^{(1)} \,\,\, = \,\,\,
D_y\,\, \,+2 \,\cdot \,{\frac {16\,{z}^{2}{y}^{2}-{z}^{3} +{y}^{2}+8\,{y}^{2}z}{
 \, (16\,{z}^{2}y+y+8\,zy-4\,{z}^{2}) \cdot  \, (4\,y-z) \cdot \,  y}}
\end{eqnarray}

\begin{eqnarray}
\hspace{-0.95in}&& \qquad
N_1^{(2)} \,\,\, = \,\,\,
 D_y\, \,\, +2\,\cdot \,{\frac {q_0(z, y)}{y \cdot \, (16\,{z}^{2}y+y+8\,zy-4\,{z}^{2}) 
\, \, (16\,y+1)  \, (z-4\,y) }}, 
\end{eqnarray}
where
\begin{eqnarray}
\hspace{-0.95in}&& \quad \,
q_0(z, y)\, \,\,=\,\,\,
-2\,{z}^{3}\,+zy\,+24\,{z}^{2}y\,+16\,{z}^{3}y\,-6\,{y}^{2}\,-40\,{y}^{2}z
\,+96\,{z}^{2}{y}^{2}\, 
 \nonumber \\
\hspace{-0.95in}&& \qquad \qquad \quad
 +128\,{z}^{3}{y}^{2}\,  -64\,{y}^{3}\,-512\,z{y}^{3}\,-1024\,{z}^{2}{y}^{3}, 
\end{eqnarray}
\begin{eqnarray}
\hspace{-0.95in}&& \qquad
N_1^{(3)}\, \,=\,\, \,\,\, N_1^{(2)}\,\,\, -{\frac {1}{ (16\,y+1) \cdot \,  y}}, 
\end{eqnarray}
and
\begin{eqnarray}
\hspace{-0.95in}&&  \qquad \qquad \quad \quad
N_2 \,\,=\,\,\,\,
 D_y^2\,\, \,\, + {\frac{\tilde{p}_1(z,y)}{\tilde{p}_2(z,y)}} \cdot \, D_y \, 
\,\,+ {\frac{\tilde{p}_0(z,y)}{\tilde{p}_2(z,y)}}, 
\end{eqnarray}
where:
\begin{eqnarray}
\hspace{-0.95in}&&  \quad
\tilde{p}_2(z,y)\, \,\,=\,\,\, \,
y \cdot \, (16\,y-1) \cdot \, (z-4\,y)
  \cdot  \, (y+8\,zy+16\,{z}^{2}y-4\,{z}^{2})
  \cdot \, (32\,{z}^{4}y
 \nonumber \\
\hspace{-0.95in}&& \quad  \quad \quad \quad \quad \quad 
-8\,{z}^{4} +256\,{z}^{3}{y}^{2} -32\,{z}^{2}{y}^{2}
+10\,{z}^{2}y -32\,{y}^{2}z+zy-2\,{y}^{2}), 
\end{eqnarray}
\begin{eqnarray}
\hspace{-0.95in}&& \quad 
\tilde{p}_1(z,y)\, \,=\,\,\,
-24\,{y}^{4}\,+15360\,{y}^{5}z\,+102400\,{y}^{5}{z}^{2}
-1310720\,{y}^{5}{z}^{5}\,-491520\,{y}^{5}{z}^{4}
 \nonumber \\
\hspace{-0.95in}&& \quad \quad \,\,  \, 
+163840\,{y}^{5}{z}^{3}\,-672\,{z}^{4}{y}^{2}\,-60\,{z}^{3}{y}^{2}
-2\,{z}^{2}{y}^{2}\,+24\,y{z}^{5}\,-3456\,{z}^{5}{y}^{2}
 \nonumber \\
\hspace{-0.95in}&& \quad  \quad  \,\,  \, 
-20480\,{z}^{6}{y}^{2}\,+576\,y{z}^{6}\,+1920\,{z}^{7}y\,-104448\,{z}^{4}{y}^{4}
+41472\,{z}^{5}{y}^{3} 
\nonumber \\
\hspace{-0.95in}&& \quad  \quad  \,\,  \, 
+32768\,{z}^{7}{y}^{3}\,+94208\,{z}^{6}{y}^{3}\,+221184\,{z}^{5}{y}^{4}
-82944\,{z}^{3}{y}^{4}\,+26112\,{z}^{4}{y}^{3} 
\nonumber \\
\hspace{-0.95in}&& \quad  \quad  \,\,  \, 
-15360\,{z}^{7}{y}^{2}\,+432\,{z}^{2}{y}^{3}\,+4672\,{z}^{3}{y}^{3}
+18\,z{y}^{3}\,-14592\,{z}^{2}{y}^{4}
 \nonumber \\
\hspace{-0.95in}&& \quad  \quad  \,\,  \, 
-1056\,z{y}^{4}\,-32\,{z}^{7}\,+640\,{y}^{5}, 
\end{eqnarray}
\begin{eqnarray}
\hspace{-0.95in}&& \quad 
\tilde{p}_0(z,y) \,=\,
-8\,{y}^{3}\,+480\,{y}^{4}\,+96\,{z}^{6}\,+10240\,{z}^{4}{y}^{2}\,-104\,{z}^{4}y
+1320\,{z}^{3}{y}^{2}
 \nonumber \\
\hspace{-0.95in}&& \quad  \quad 
+152\,{z}^{2}{y}^{2}\,+8\,{y}^{2}z\,-4\,{z}^{3}y\,-672\,y{z}^{5}\,+34048\,{z}^{5}{y}^{2}
+104448\,{z}^{6}{y}^{2}
\nonumber \\
\hspace{-0.95in}&& \quad  \quad 
-11648\,y{z}^{6} \,-8704\,{z}^{7}y\,-368640\,{z}^{4}{y}^{4}\,+2048\,{z}^{5}{y}^{3}-155648
\,{z}^{6}{y}^{3}
 \nonumber \\
\hspace{-0.95in}&& \quad  \quad 
-983040\,{z}^{5}{y}^{4}\,+122880\,{z}^{3}{y}^{4}\,
-64512\,{z}^{4}{y}^{3}\,+18432\,{z}^{7}{y}^{2}\,-8544\,{z}^{2}{y}^{3}
 \nonumber \\
\hspace{-0.95in}&& \quad  \quad 
-44288\,{z}^{3}{y}^{3} \,-568\,z{y}^{3}\,+76800\,{z}^{2}{y}^{4}\,
+11520\,z{y}^{4}\,-8\,{z}^{5}\,+640\,{z}^{7}.
\end{eqnarray}

\section{The matching of $\, T_2(z, y)$ with the solutions of $ \, L_5^{(z)}$}
\label{combODEzT2}

$T_2(z,y)$ as a linear combination on the formal solutions of $ \, L_5^{(z)}$ reads
\begin{eqnarray}
\label{T2zy}
\hspace{-0.95in}&& \qquad  \quad \,
T_2(z, y)\, \,=\,\,\,\,
 C_1^{(z)}(y) \cdot S_1^{(z)}\, \,+ C_2^{(z)}(y) \cdot S_2^{(z)}\, \,+ 
C_3^{(z)}(y) \cdot S_3^{(z)}, 
\end{eqnarray}
where
\begin{eqnarray}
\label{S1z}
\hspace{-0.95in}&& \quad  \quad 
S_1^{(z)}\, \,=\,\,\,
1\,\, + \left(-{16 \over 3}+2\,{y}^{-1} \right) \cdot \, {z}^{2}\,
\,+ \left( {\frac {512}{15}}-{\frac {184}{15}}\,{y}^{-1}\right) \cdot \,{z}^{3}
\nonumber \\
\hspace{-0.95in}&& \qquad  \quad  \quad  \quad 
  + \Bigl( -{\frac {18176}{105}}+
{\frac {320}{7}}\,{y}^{-1}  
+{\frac {208}{35}}\,{y}^{-2} \Bigr) \cdot \, {z}^{4}
\nonumber \\
\hspace{-0.95in}&& \qquad  \quad  \quad  \quad 
+ \left( {\frac {253952}{315}}-{\frac {2304}{35}}\,{y}^{-1}
-{\frac {2816}{35}}\,{y}^{-2}-{\frac {4}{315}}\,{y}^{-3} \right)
\cdot \, {z}^{5} \, \, \,
+ \,  \cdots 
\nonumber
\end{eqnarray}
\begin{eqnarray}
\hspace{-0.95in}&& \quad  \quad  
S_2^{(z)} \,\,=\,\,\,\,
z \, \,\, 
+ \left( -{16 \over 3}+ {1 \over 6}\,{y}^{-1} \right)\cdot \,  {z}^{2}\, \,
+ \left( {\frac {368}{15}}+{\frac {22}{15}}\,{y}^{-1}
+{1 \over 30}\,{y}^{-2} \right) \cdot \,  {z}^{3}
\nonumber \\ 
 \hspace{-0.95in}&& \quad  \quad  \quad   \quad   \quad   \quad  
+\left( -{\frac {11264}{105}}
-{\frac {2864}{105}}\,{y}^{-1}+{\frac {8}{35}}\,{y}^{-2}
+{\frac {1}{140}}\,{y}^{-3} \right) \cdot \, {z}^{4}\,
 \nonumber \\
 \hspace{-0.95in}&& \quad  \quad  \quad   \quad   \quad   \quad  
+ \Bigl( {\frac {144128}{315}}+{\frac {79424}{315}}\,{y}^{-1}
+{\frac {96}{35}}\,{y}^{-2}+{\frac {2}{45}}\,{y}^{-3}
+{\frac {1}{630}}\,{y}^{-4} \Bigr) \cdot \, {z}^{5}\, \,\,  + \, \cdots 
\nonumber
\end{eqnarray}
\begin{eqnarray}
\hspace{-0.95in}&& \quad  \quad  
S_3^{(z)} \,\,=\,\,\,\,
{z}^{2} \, \,\, -6\,{z}^{3} \, \,+ \left( 36+3\,{y}^{-1} \right) \cdot  \, {z}^{4}\,
 \,- \left(180+40\,{y}^{-1} \right) \cdot \,{z}^{5}   
\nonumber \\
\hspace{-0.95in}&& \quad  \quad  \quad   \quad  \quad   \quad  
+\left( 900+370\,{y}^{-1} +10\,{y}^{-2} \right)\cdot \,  {z}^{6}
 \, \, \,  + \, \cdots
 \nonumber 
\end{eqnarray}
\begin{eqnarray}
\hspace{-0.95in}&& \quad  \quad  
C_1^{(z)}(y) \,\,=\,\,\,
{\frac {1}{\sqrt {1-16\,y}}} \, \cdot \,
_{2}F_1 \left( [{{1}  \over {2}}, \, {{1}  \over {2}}],[1], \, 
16 \cdot \,{\frac {y}{16\,y-1}} \right),  
\end{eqnarray}
\begin{eqnarray}
\hspace{-0.95in}&& \quad  \quad  
C_2^{(z)}(y) \,\,=\,\,\,
-{\frac{8\,y}{(8\,y-1)^{3/2}}} \, \cdot \,
_{2}F_1\left( [{{3}  \over {4}}, \,{{5}  \over {4}}],[2], \, 
64 \cdot \,{\frac {{y}^{2}}{ (1-8\,y)^{2}}} \right), 
\end{eqnarray}
\begin{eqnarray}
\hspace{-0.95in}&& \quad  \quad 
C_3^{(z)}(y) \,\,=\,\,\, -2\,{y}^{-1}. 
\end{eqnarray}

\section{The matching of $\,T_2(z, y)$ with the solutions of $\,N_5^{(y)}$}
\label{combODEyT2}

$T_2(z,y)$ as linear combination on the formal solutions of $\, N_5^{(y)}$ reads
\begin{eqnarray}
\hspace{-0.95in}&& \quad  \quad 
T_2(z, y) \, \,=\,\,  \,\, C_1^{(y)}(z) \cdot S_1^{(y)} \, \, + C_2^{(y)}(z) \cdot S_2^{(y)}
\, \,  + C_3^{(y)}(z) \cdot S_3^{(y)}, 
\end{eqnarray}
where
\begin{eqnarray}
 \hspace{-0.95in}&& \quad  \quad 
S_1^{(y)}\,  \,=\,\, \, \,
1\,\, \, -16\,y\, \, \, - \left({\frac {128}{3}} +{\frac {208}{3}}\,{z}^{-1}
+{16 \over 3}\,{z}^{-2} +{1 \over 3}\,{z}^{-3} \right)\cdot \,  {y}^{2}
 \nonumber \\
 \hspace{-0.95in}&& \quad  \quad  \quad \quad  \quad 
-\left({\frac {2048}{15}}\,
+{\frac {4096}{15}}\,{z}^{-1} +{\frac {1408}{5}}\,{z}^{-2}
+{\frac {464}{15}}\,{z}^{-3} +{8 \over 3}\,{z}^{-4} +{1 \over 15}\,{z}^{-5} \right)
 \cdot {y}^{3} \,  + \,\cdots 
\nonumber
\end{eqnarray}
\begin{eqnarray}
\hspace{-0.95in}&& \quad  \quad 
S_2^{(y)}\, \,=\,\,\, \,
y \,\,  \,
+ \left( {8 \over 3}+4\,{z}^{-1}+{1 \over 6}\,{z}^{-2} \right)\cdot \,  {y}^{2}
 \nonumber \\
 \hspace{-0.95in}&& \quad  \quad  \quad \quad  \quad 
+ \left( {
\frac {128}{15}}+16\,{z}^{-1}
+{\frac {232}{15}}\,{z}^{-2}+{z}^{-3}+{1 \over 30}\,{z}^{-4} \right) \cdot \, {y}^{3}
 \,\, \, + \,\,\cdots
\nonumber 
\end{eqnarray}
\begin{eqnarray}
\hspace{-0.95in}&& \quad  \quad 
S_3^{(y)}\, \,=\,\,\,\, \,
y\,\, \, \, + \left( 8+3\,{z}^{-1} \right) \cdot \, {y}^{2} 
 \,\,\, + \left( 76+30\,{z}^{-1}+10\,{z}^{-2} \right) \cdot \, {y}^{3} 
\,\, \, + \,\,\cdots 
\nonumber
\end{eqnarray}
\begin{eqnarray}
\hspace{-0.95in}&& \quad  \quad 
C_1^{(y)}(z) \,\,=\,\,\,\,\,
{\frac{1}{\sqrt{1-16 z^2}}} \, \cdot \,
_{2}F_1\left( [{{1}  \over {2}}, \, {{1}  \over {2}}],[1],{\frac {16 z^2}{16 z^2-1}} \right),   
\end{eqnarray}
\begin{eqnarray}
\hspace{-0.95in}&& \quad  \quad 
C_2^{(y)}(z) \,\,=\,\,\,\,\,
{\frac { 2 \cdot \, \, \left( 12\,z+1 \right) }{z}}\cdot \,  H_z(z) \,  \,  \, 
+{\frac { \left( 4\,z+1 \right)  \left( 4\,z-1 \right) }{2 z}} \cdot \, 
 {\frac{d}{dz}} H_z(z), \\
\hspace{-0.95in}&& \quad  \quad  
\hbox{with:}  \quad  \quad  \quad \quad  \quad  H_z(z)\,  \,=\, \,\,
 _{2}F_1\left( [{{1}  \over {2}}, \, {{1}  \over {2}}],[1], \, 16\, z^2 \right), 
\end{eqnarray}
\begin{eqnarray}
\hspace{-0.95in}&& \quad  \quad 
C_3^{(y)}(z) \,\,=\,\,\, -2\,{z}^{-1}.
\end{eqnarray}

\section{Closed form expression of $\, V_2(z,y)$}
\label{V2}

The bivariate series
\begin{eqnarray}
\hspace{-0.95in}&&   \quad  \qquad  \quad   \quad 
V_2 (z, y) \,\,=\,\,\,\, 
{\frac{1}{\pi^2}} \,\, \int_0^\pi  \int_0^\pi \,\,
 {\frac{\mathrm{d}k_1\, \mathrm{d}k_2}{(1\,- z \, \zeta_2) 
\cdot \, (1\,- {y \over 2} \, \sigma_2)}}, 
\end{eqnarray}
with
\begin{eqnarray}
\hspace{-0.95in}&&   \quad  \qquad  \quad   \quad 
\zeta_2 \,=\,\,  \cos(k_1) \cdot \, \cos(k_2), \qquad \quad 
\sigma_2 \,=\,\,  \cos(k_1)\,  + \cos(k_2), 
\end{eqnarray}
is related to the integral (\ref{integralTdzy}) with $d= \, 2$ by 
$T_2(z,y) \,=\, V_2(4 z, \pm 4 \sqrt{y})$.
The integral $\, V_2(z, y)$ can be written in the closed form expression
\begin{eqnarray}
\hspace{-0.95in}&&     \    \quad 
V_2^{closed}(z,y)\,  \, =\, \,\, \, \, 
 {\frac{y}{z y + y -2 z }} \cdot \, K(y) \,  \,  \,\,  
 +  {\frac{z}{ \delta }} \cdot \, \Bigl(\Pi ({\frac{(z-\delta)^2}{y^2}}, z)\,  \, 
 -\Pi ({\frac{(z+\delta)^2}{y^2}}, z) \Bigr)
 \nonumber \\
\hspace{-0.95in}&&   \quad  \quad \quad   \qquad
+ {\frac{y \cdot \, (1-y) \cdot \, \delta}{(z-y^2) (z y + y\, -2 z)}} \cdot \, 
\Bigl(\Pi (\Delta_{-},\, y ) \, - \Pi (\Delta_{+},\, y ) \Bigr), 
\end{eqnarray}
with 
\begin{eqnarray}
\hspace{-0.95in}&&     \,    \quad    \quad   
\delta\,\,   = \,\, \,  \sqrt{z^2 -z y^2},
 \quad \quad \,\, \, 
\Delta_{\pm}\,  = \,\, \, 
 {\frac{ 2 z y+y^2 -z y^2 \, 
-2 z \,\, \pm   2\,  (1-y) \cdot \, \delta }{z y + y -2 z }}, 
\end{eqnarray}
and where $ \, \Pi$ and $ \, K$ are the 
complete elliptic integrals of the third and first kind:
\begin{eqnarray}
\hspace{-0.95in}&&       
\Pi( \nu, k)  \,=\,\,  \int_0^{\pi} \, {\frac{1}{(1-\nu \cos(\phi)^2)}} \cdot \, 
{\frac{1}{\sqrt{1-k^2 \cos(\phi)^2}}} \cdot \, \mathrm{d}\phi, 
\quad \, \, \, 
 K(k) \, = \, \,\,  \Pi(0, k).
\end{eqnarray}
One has
\begin{eqnarray}
 \hspace{-0.75in}&&  \quad      \quad      
V_2(z,y) \,\,  =\,\,\,   Re \left( V_2^{closed}(z,y) \right), 
\quad  \quad  \quad      \,\, 
{\rm for} \,\,  \,\, \vert z \vert <1, \, \, \,\,  \vert y \vert <1, 
\end{eqnarray}
and
\begin{eqnarray}
\hspace{-0.75in}&&  \quad    \quad             
V_2(z,y) \, \, =\, \, \,  V_2^{closed}(z,y),
 \quad \quad \quad      \,\, 
{\rm for} \quad  \,\,  y \cdot \, (z y +y -2 z) \, > \, 0. 
\end{eqnarray}

\section{The two right-most order-two operators 
            $\, L_2$ and $\,N_2$ for $\, T_3(z,y)$}
\label{rightmost}

$\, \bullet$ The  right-most order-two linear differential operator
$\, L_2$  (see (\ref{L9z}))
in the factor\-ization of the order-nine operators $\, L_9^{(z)}$ 
reads:
\begin{eqnarray}
\hspace{-0.95in}&&  \quad  \quad \quad \quad \quad \quad 
L_2 \,\,=\,\,\,\,  D_z^2 \,\,\,\,
+ z^2 \cdot \, {\frac{p_1(z,y)}{p_2(z,y)}} \cdot \, D_z 
\,\,\, \,+ {\frac{p_0(z,y)}{p_2(z,y)}}, 
\end{eqnarray}
where:
\begin{eqnarray}
\hspace{-0.95in}&&  \quad  
p_2(z,y) \,\,=\,\,\,\,  {z}^{2} \cdot \,
\left( 432\,{y}^{2}{z}^{2}+180\,z{y}^{2}-36\,{z}^{2}y+12\,{y}^{2}-13\,yz+2\,{z}^{2} \right) 
\nonumber \\
\hspace{-0.95in}&& \quad \quad \quad   \quad  \times 
(4\,z+1) \cdot \, (3\,y-z) \cdot \, 
(144\,{z}^{2}y+24\,yz-16\,{z}^{2}+y) 
 \nonumber \\
\hspace{-0.95in}&& \quad \quad \quad \quad   \times 
(144\,{y}^{2}{z}^{2}+96\,z{y}^{2}-40\,{z}^{2}y+16\,{y}^{2}-8\,yz+{z}^{2}), 
\end{eqnarray}
\begin{eqnarray}
\hspace{-0.95in}&&    \quad  
p_1(z,y) \,\,=\,\,\,\,
512\, \cdot \, (4\,y-1)  \cdot \, (36\,y-1)  \cdot \, (9\,y-1) 
 \left( 216\,{y}^{2}-18\,y+1 \right) \cdot \,  {z}^{7}
\nonumber \\
\hspace{-0.95in}&& \quad \quad 
\, -32\, \cdot \, (10077696\,{y}^{6}-10730880\,{y}^{5}
+3050784\,{y}^{4}
-361080\,{y}^{3}+20176\,{y}^{2}-481\,y+3) \cdot \,   {z}^{6}
\nonumber \\
\hspace{-0.95in}&& \quad \quad   
\, -16\,y \left( 25754112\,{y}^{5}-16422912\,{y}^{4}
+3259008\,{y}^{3}-270672\,{y}^{2}+9518\,y-103
 \right)\cdot \,  {z}^{5}
\nonumber \\
\hspace{-0.95in}&& \quad \quad   
-2\,y \left( 103762944\,{y}^{5}-46033920\,{y}^{4}
+6379200\,{y}^{3}-334704\,{y}^{2}+5464\,y-1 \right)\cdot \,   {z}^{4} \,
 \nonumber \\
\hspace{-0.95in}&& \quad \quad  
-2\,{y}^{2} \cdot \, (26065152\,{y}^{4}-
8142336\,{y}^{3}+717600\,{y}^{2}-17920\,y+13)\cdot \,   {z}^{3}
\nonumber \\
\hspace{-0.95in}&& \quad \quad    
\, -{y}^{3} \, 
(6816960\,{y}^{3}-1407888\,{y}^{2}+60196\,y-99) \cdot \,   {z}^{2}
\nonumber \\
\hspace{-0.95in}&& \quad \quad 
\, -24\,{y}^{4} \, (18288\,{y}^{2}-1960\,y+3) \cdot \,   z\, \, 
-10944\,{y}^{6}-144\,{y}^{5}, 
\end{eqnarray}
\begin{eqnarray}
\hspace{-0.95in}&&    \quad  
p_0(z,y) \,\,=\,\,\,\,
256\, \, (4\,y-1)  \, (36\,y-1)  \, (9\,y-1) 
 \, (216\,{y}^{2}-18\,y+1) \cdot \, {z}^{8}
\nonumber \\
\hspace{-0.95in}&& \quad \quad   
\, - \, (107495424\,{y}^{6}-134369280\,{y}^{5}+41720832\,{y}^{4}
-5163264\,{y}^{3}+284928\,{y}^{2}-6240\,y+32) \cdot \,  {z}^{7}
\nonumber \\
\hspace{-0.95in}&& \quad \quad   
-8\,y \, (14556672\,{y}^{5}-10917504\,{y}^{4}+2372544\,{y}^{3}
-202944\,{y}^{2}+6788\,y-63) \cdot   \,  {z}^{6}
\nonumber \\
\hspace{-0.95in}&& \quad \quad  
\, -16\, (3032640\,{y}^{4}-1574640\,{y}^{3}+235404\,{y}^{2}
-12305\,y+178) {y}^{2}\cdot \, {z}^{5}
\nonumber \\
\hspace{-0.95in}&& \quad \quad    
\, -4\,{y}^{2} \, (2519424\,{y}^{4}-883872\,{y}^{3}
+81840\,{y}^{2}-1956\,y+1) \cdot \,  {z}^{4} \, 
\nonumber \\
\hspace{-0.95in}&& \quad \quad  
-16\,{y}^{3} \cdot \, (75168\,{y}^{3}-15768\,{y}^{2}+746\,y-3) \cdot \,  {z}^{3}
\nonumber \\
\hspace{-0.95in}&& \quad \quad   
\, -2\, \cdot \, (49032\,{y}^{2}-5550\,y+113) \cdot \, {y}^{4} \cdot \, {z}^{2} 
\,\,  -12\, \cdot \, (456\,y-35) {y}^{5} \cdot \, z \, 
-144\,{y}^{6}, 
\end{eqnarray}

This order-two operator $\, L_2$ is self-adjoint up to a 
conjugation by its Wronskian  $\, W(L_2)$
\begin{eqnarray}
\hspace{-0.95in}&& \quad \quad   \qquad
L_2 \cdot \, W(L_2)  \,  \, =\,\,  \, W(L_2)  \cdot \, adjoint(L_2), \qquad 
\end{eqnarray}
where this Wronskian  $\, W(L_2)$  reads:
\begin{eqnarray}
\label{WL2}
\hspace{-0.95in}&&  \quad 
W(L_2) \, \,= \, \, \,
{\frac {432\,{y}^{2}{z}^{2}+180\,{y}^{2}z-36\,y{z}^{2}+12\,{y}^{2}-13\,yz+2\,{z}^{2}
}{(4\, z\, +1) \cdot \, (z \,-3\,y) \cdot  \, d_1 
 \cdot \, d_2 }} \qquad 
 \\
\hspace{-0.95in}&& \qquad \quad   \hbox{where:}   \qquad   \quad 
 d_1 \, \, = \, \, \, 144\,y{z}^{2}+24\,yz-16\,{z}^{2}+y, 
 \nonumber \\
\hspace{-0.95in}&&  \qquad \quad  \hbox{and:}  \qquad   \quad \quad  
d_2 \, \, = \, \, \,
144\,{y}^{2}{z}^{2}+96\,{y}^{2}z-40\,y{z}^{2}+16\,{y}^{2}-8\,yz+{z}^{2}. 
\nonumber
\end{eqnarray}

$\, \bullet$ The  right-most order-two linear differential operator 
 $\, N_2$ (see (\ref{N9y}))
in the factorization of the order-nine operators  $\, N_9^{y}$ 
reads:
\begin{eqnarray}
\hspace{-0.95in}&&  \quad  \quad \quad \quad \quad \quad 
N_2 \,\,=\,\,\,\, D_y^2 \,\,\,\,
+ y^2 \cdot \, {\frac{q_1(z,y)}{q_2(z,y)}} \cdot \, D_y \,\,\, \,
+ {\frac{ q_0(z,y)}{q_2(z,y)}}, 
\end{eqnarray}
where:
\begin{eqnarray}
\hspace{-0.95in}&&   \quad \quad 
q_2(z,y) \,\,=\,\,\,
{y}^{2} \cdot \, (36\,yz-6\,y+z) \cdot \, (3\,y-z)
 \nonumber \\
\hspace{-0.95in}&& \quad \quad \quad \quad \quad \times 
 (144\,{y}^{2}{z}^{2}+96\,{y}^{2}z-40\,y{z}^{2}+16\,{y}^{2}-8\,yz+{z}^{2})
\nonumber \\
 \hspace{-0.95in}&& \quad \quad \quad \quad \quad \times 
   (144\,y{z}^{2}+24\,yz-16\,{z}^{2}+y), 
\end{eqnarray}
\begin{eqnarray}
\hspace{-0.95in}&&   \quad \quad 
q_1(z,y) \,\,=\,\,\,
864\, \cdot \, (6\,z-1)  \cdot \,(3\,z+1)^{2} \cdot \, (12\,z+1)^{2} \cdot \, {y}^{4}
\nonumber \\
\hspace{-0.95in}&& \quad \quad \quad 
\, -96\,z \cdot \, 
(15552\,{z}^{5}+25920\,{z}^{4}+4428\,{z}^{3}-1098\,{z}^{2}-273\,z-7) \cdot \, {y}^{3}
\nonumber \\
\hspace{-0.95in}&& \quad \quad \quad 
\, +6\,{z}^{2} \cdot \,
 (38016\,{z}^{4}+7200\,{z}^{3}-8184\,{z}^{2}-1850\,z-31) \cdot \,  {y}^{2}
\\
\hspace{-0.95in}&& \quad \quad \quad 
\, +2\,{z}^{3} \cdot \, 
(8064\,{z}^{3}+6000\,{z}^{2}+944\,z+11) \cdot \,  y \,
\, \,  -1360\,{z}^{6}-104\,{z}^{5}-{z}^{4}, 
\nonumber 
\end{eqnarray}
\begin{eqnarray}
\hspace{-0.95in}&&   \quad \quad 
q_0(z,y) \,\,=\,\,\,\, 
4\,{z}^{6} \,\, +16\, \cdot \, (25\,z-4) \cdot \,  {z}^{5} \cdot \, y \,\, 
\, \, 
\nonumber \\
\hspace{-0.95in}&& \quad \quad \quad 
-2\,{z}^{3} \cdot \,
 (2448\,{z}^{3}+24\,{z}^{2}-163\,z-1) \cdot \,  {y}^{2}
\nonumber \\
\hspace{-0.95in}&& \quad \quad \quad 
\, +12\,{z}^{2} \cdot \, 
(432\,{z}^{4}-792\,{z}^{3}-669\,{z}^{2}-113\,z-2) \cdot \,  {y}^{3} \,
 \nonumber \\
\hspace{-0.95in}&& \quad \quad \quad 
-12\,z \cdot \, 
(15552\,{z}^{5}+22032\,{z}^{4}+1188\,{z}^{3}-1989\,{z}^{2}-363\,z-10)\cdot \,  {y}^{4}
\nonumber \\
\hspace{-0.95in}&& \quad \quad \quad 
\, +216\, \cdot \, (6\,z-1)  \, (3\,z+1)^{2} \, (12\,z+1)^{2} \cdot \, {y}^{5}. 
\end{eqnarray}
This order-two operator $\, N_2$ is self-adjoint up to a 
conjugation by its Wronskian  $\, W(N_2)$
\begin{eqnarray}
\hspace{-0.95in}&& \quad \quad   \qquad
N_2 \cdot \, W(N_2)  \,  \, =\,\,  \, W(N_2)  \cdot \, adjoint(N_2), \qquad 
\end{eqnarray}
where this Wronskian  $\, W(N_2)$  reads:
\begin{eqnarray}
\hspace{-0.95in}&& \quad  
W(N_2) \, \,= \, \, \,
{\frac {36\,yz-6\,y+z}{z \cdot \, (z \, -3\,y) \cdot  \, d_1 
 \cdot \, d_2 }} \quad \hbox{where:}  \quad \quad 
 d_1 \, \, = \, \, \, 144\,y{z}^{2}+24\,yz-16\,{z}^{2}+y, 
 \nonumber \\
\hspace{-0.95in}&& \hbox{and:}  \qquad \quad 
d_2 \, \, = \, \, \,
144\,{y}^{2}{z}^{2}+96\,{y}^{2}z-40\,y{z}^{2}+16\,{y}^{2}-8\,yz+{z}^{2}.
\end{eqnarray}

\vskip .1cm 

\section{Singularities of the $\, T_d(z,y)$}
\label{singTdzy}

Based on the singularities of the ODEs in one variable (the other being
a parameter) or Landau conditions methods~\cite{2007-PhiH-integrals}, 
we have the following results.
The bivariate series have singularities bearing on the variable $\,z$, 
these singularities are those of the linear ODEs of the LGF of the $d$-dimensional 
fcc lattice. Similarly, the singularities corresponding to the simple 
lattice appear as singularities in the variable $\,y$.
Besides these obvious and expected singularities, one obtains algebraic curves
on $\, (z,\,  y)$ as singular varieties. For $d=\, 2$, $d=\,3$ and $d=\,4$, 
they read:
\begin{eqnarray}
\hspace{-0.95in}&&   \quad \quad \quad 
d = 2, \quad \quad 4 y -z\,\,  =\,\, 0, \quad \quad\quad 
 4\,{z}^{2} \cdot \, (4\,y-1)\, +8\,yz+y\,  \,= \,\,\,  0, 
\nonumber \\
\hspace{-0.95in}&&   \quad  \quad \quad 
d = 3, \quad \quad 3 y -z\,\,  =\,\, 0, \quad \quad \, \, \,
 16\,{z}^{2} \cdot \, (9\,y-1) \,+24\,yz+y \, \, =\,\,\, 0,
 \nonumber \\
\hspace{-0.95in}&&   \quad  \qquad \qquad \quad \, \, \, \, \,
16\,{y}^{2} \cdot \, (3\,z+1)^{2}\,
 -z \cdot \, (40\,yz+8\,y-z) \, \, =\,\, \,0,  \\
\hspace{-0.95in}&&   \quad  \quad \quad 
d = 4, \quad \quad  8\,y-3\,z \, \,=\,\,0, \quad \quad \quad 
36\,{z}^{2} \cdot  \, (16\,y-1)\, +48\,yz+y
\,\,  =\,\, \,0, 
\nonumber \\
\hspace{-0.95in}&&   \quad  \quad \qquad \qquad \, \, \,\, \,
4\,{z}^{2} \cdot \, (16\,y-1) \,+16\,yz +y \, \,=\,\,\,  0, 
\quad \quad 
16\,yz +4\,y-z \,= \,\,0.
\nonumber 
\end{eqnarray}

One may infer from the first two varieties of each $d$, the expressions
in function of the dimension
\begin{eqnarray}
\hspace{-0.95in}&&   \quad  \qquad 
y \,\,\,  - {\frac{d-1}{2 d}} \cdot \, z \, \, \,= \,\,\,\,  0, 
\quad \quad  \quad  
y \, \, \, - {\frac {4 \cdot \, (d-1)^{2} \cdot \,{z}^{2} }{
 ( 1 \,+2 \cdot \, d \cdot \, (d-1) \cdot \,z)^{2}}} \,\, \, = \,\,\,  0.
\end{eqnarray}

\end{document}